\documentclass[journal]{IEEEtran}
\usepackage{amsmath,amsfonts}
\usepackage{algorithmic}
\usepackage{algorithm}
\usepackage{amssymb}
\usepackage{multirow}
\usepackage{array}
\usepackage{subfig}
\usepackage{textcomp}
\usepackage{stfloats}
\usepackage{url}
\usepackage{verbatim}
\usepackage{graphicx}
\usepackage{cite}
\usepackage{hyperref}
\newcommand{\normalize}[1]{\ensuremath{#1}}

\hypersetup{
    colorlinks=true,
    linkcolor=blue,
    filecolor=magenta,      
    urlcolor=cyan,
    citecolor = magenta,
    pdfpagemode=FullScreen,
}

\begin{document}

\title{\fontsize{20.0}{23.4}\selectfont{Cross-Layer Design of Vector-Symbolic Computing: Bridging Cognition and Brain-Inspired Hardware Acceleration}}

\author{
Shuting Du,
Mohamed Ibrahim,
Zishen Wan,
Luqi Zheng,
Boheng Zhao,
Zhenkun Fan,
Che-Kai Liu,
Tushar Krishna,
Arijit Raychowdhury,
and Haitong Li

\thanks{Shuting Du, Luqi Zheng, Boheng Zhao, and Haitong Li are with the Elmore Family School of Electrical and Computer Engineering, Purdue University, West Lafayette, IN 47907 USA (e-mail: du335@purdue.edu; haitongli@purdue.edu).}
\thanks{Mohamed Ibrahim, Zishen Wan, Zhenkun Fan, Che-Kai Liu, Tushar Krishna, and Arijit Raychowdhury are with the School of Electrical and Computer Engineering, Georgia Institute of Technology, Atlanta, GA 30332 USA (e-mail: mohamed.ibrahim@gatech.edu; arijit.raychowdhury@ece.gatech.edu).}
       
\thanks{This work was supported in part by the U.S. National Science Foundation under Award No. 2425498 with industry partners as specified in the Future of Semiconductors (FuSe2) program, and in part by the UPWARDS for the Future Network program (with funding from NSF Award No. 2329784, Micron Technology and TEL). The work was also supported in part by CoCoSys, one of seven centers in JUMP 2.0, a Semiconductor Research Corporation (SRC) program sponsored by DARPA. \textit{(Shuting Du and Mohamed Ibrahim contributed equally to this work. Corresponding authors: Shuting Du, Mohamed Ibrahim, Arijit Raychowdhury, Haitong Li.)}}
}

\maketitle

\begin{abstract}

Vector Symbolic Architectures (VSAs) have been widely deployed in various cognitive applications due to their simple and efficient operations. The widespread adoption of VSAs has, in turn, spurred the development of numerous hardware solutions aimed at optimizing their performance. Despite these advancements, a comprehensive and unified discourse on the convergence of hardware and algorithms in the context of VSAs remains somewhat limited. The paper aims to bridge the gap between theoretical software-level explorations and the development of efficient hardware architectures and emerging technology fabrics for VSAs, providing insights from the co-design aspect for researchers from either side. First, we introduce the principles of vector-symbolic computing, including its core mathematical operations and learning paradigms. Second, we provide an in-depth discussion on hardware technologies for VSAs, analyzing analog, mixed-signal, and digital circuit design styles. We compare hardware implementations of VSAs by carrying out detailed analysis of their performance characteristics and tradeoffs, allowing us to extract design guidelines for the development of arbitrary VSA formulations. Third, we discuss a methodology for cross-layer design of VSAs that identifies synergies across layers and explores key ingredients for hardware/software co-design of VSAs. Finally, as a concrete demonstration of this methodology, we propose the first in-memory computing hierarchical cognition hardware system, showcasing the efficiency, flexibility, and scalability of this co-design approach. The paper concludes with a discussion of open research challenges for future explorations.

\end{abstract}

\section{Introduction}
\label{sec:intro}

\IEEEPARstart{P}{rocessing} on streams of real-time data was one of the main motivators behind the early development of electronic circuits. Hence, it was no surprise that the enabling of digital integrated computing in the 1970s also led to specialized processors for signal processing. While initially focused on single-dimensional streams (e.g., audio~\cite{frantz1982design}), advances in image and video processing rapidly led to processors that operated on two-dimensional and three-dimensional data, leading to vector and matrix processors (early 1980s)~\cite{kung1982systolic}. With digital neural networks gaining popularity, the need to operate on even higher-dimensional structures such as tensors has become necessary, leading to processors such as the Google TPU~\cite{jouppi2017datacenter}. This trajectory suggests a continued trend towards computing in increasingly higher-dimensional spaces, driven by innovative applications across various domains, such as security, optimization, and cognition. Therefore, computing in high-dimensional (HD) spaces has received substantial interest not only from mathematicians but also from researchers in neuroscience and electronic-circuit design~\cite{vershynin2018high, rahimi2017high, eliasmith2012large}.

Among various approaches to computing in HD spaces, the field of \emph{Vector Symbolic Architectures} (VSAs)---also referred to as Hyperdimensional Computing (HDC)---has emerged as a promising and effective approach, taking inspiration and guidance from the neural architecture of biological systems~\cite{kanerva2009hyperdimensional}. The conceptualization of this approach is founded on the idea that key aspects of human memory, perception, and cognition can be modeled by computing with HD distributed vector representations (or ``hypervectors''), which capture the rich phenomena inherent in the collective activity of large populations of neurons in the brain~\cite{churchland2012neural, stringer2019high}. Sensory data, state variables, and high-level concepts are all mapped into vectors in an HD space, and an algebra over these vectors is then used to combine information to form new representations that can be used as the basis for further processing~\cite{kleyko2022vector}. This compositional structure is central to the expressive power of VSAs, which thereby lend themselves to supporting a wide range of cognitive tasks, including classification, factorization, planning, reasoning, and problem-solving~\cite{kleyko2023survey, menon2023shared, stockl2024local, frady2020resonator, hersche2023neuro}. 

Furthermore, VSAs exhibit three inherent qualities that naturally facilitate efficient hardware implementation. The first is \emph{robustness to errors}, a capability that stems from computing with distributed HD representations~\cite{rahimi2016robust, morris2021hydrea}. The impact of this capability is that VSA hardware does not require ultra-reliable elements, enabling operation with minimal energy consumption even under low signal-to-noise conditions~\cite{burrello2019hyperdimensional, rahimi2017high}. The second quality is \emph{element-wise independence}, which arises as computations are performed independently across all vector elements. This independence permits significant computational acceleration, with multiple operations executed simultaneously across the hardware fabric~\cite{chen2022full}. The third quality is \emph{versatility}, which enables rematerialization of VSA primitives, potentially on-the-fly, to adapt to a wide range of computational tasks~\cite{schmuck2019hardware}. This flexibility is due to the ability of HD vectors to encode diverse types of data and relationships within a unified framework. Together, these qualities lay the foundation for growing research on VSA hardware development, supported by ongoing advancements in semiconductor technologies~\cite{akarvardar2023technology} and physical design flows~\cite{liew2022hammer}.

Hence, research on the hardware development of VSAs has received significant attention in recent years, leveraging a broad diversity of hardware platforms. These include embedded CPUs/GPUs~\cite{montagna2018pulp, kang2022openhd, simon2022hdtorch}, FPGA platforms~\cite{salamat2019f5, imani2021revisiting, salamat2020accelerating}, custom application-specific integrated circuits (ASICs)~\cite{datta2023hdbinarycore, eggimann20215, ibrahim2024efficient}, neuromorphic spiking arrays (NSA)~\cite{bent2022hyperdimensional, orchard2023hyperdimensional, renner2022sparse}, and emerging non-volatile memory (NVM) fabrics~\cite{li2016hyperdimensional, karunaratne2020memory}. Studies in this field have demonstrated highly efficient solutions, achieving energy savings and performance enhancements of several orders of magnitude compared to VSA implementations on general-purpose processors. Consequently, hardware advancements in VSAs have become crucial enablers for machine intelligence at the edge, providing advanced cognitive capabilities while meeting stringent energy and form-factor constraints~\cite{chang2023recent, menon2022highly, yu2023fully}. 

\begin{figure}[!t]
	\centering
	\includegraphics[width=\columnwidth]{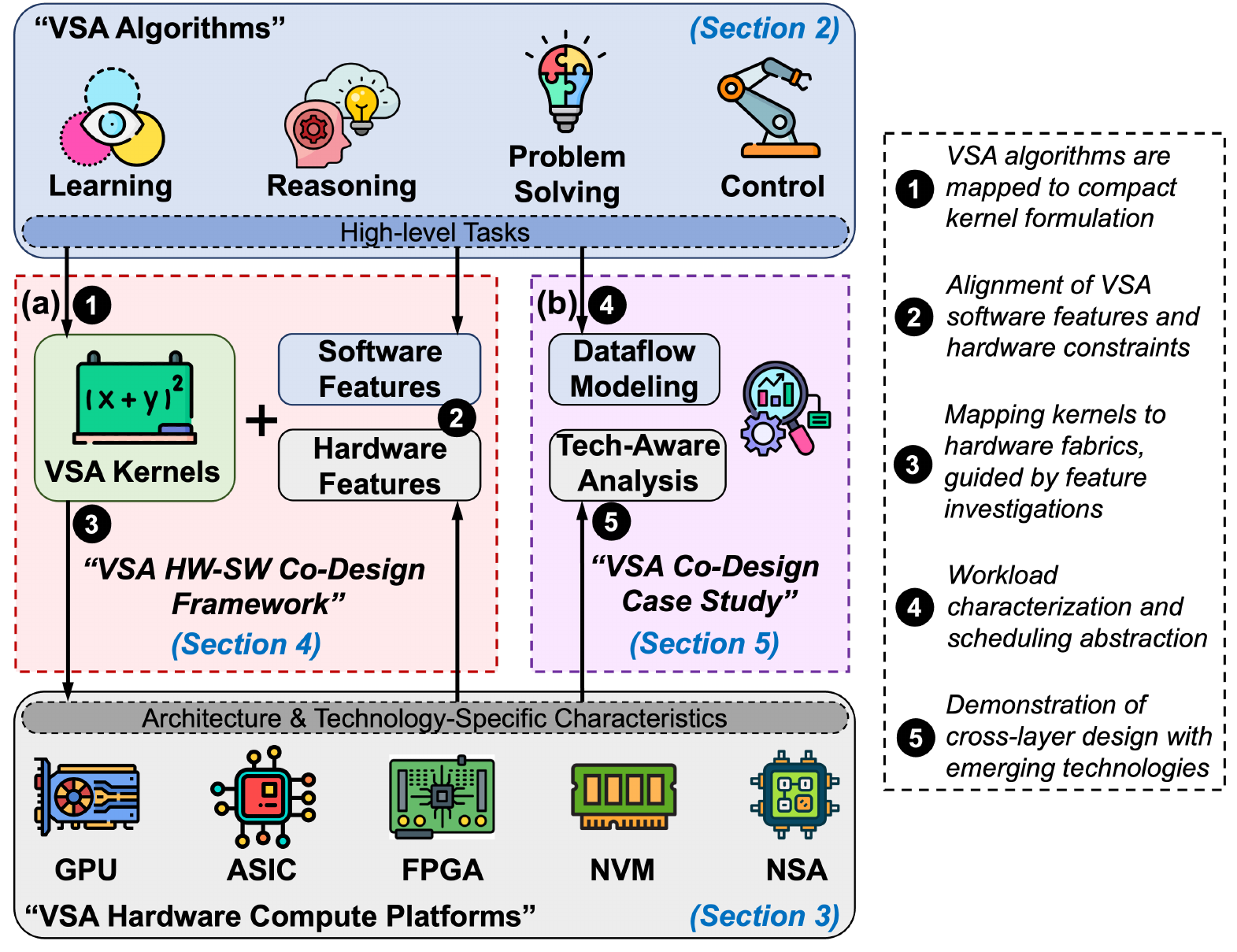}
	\caption{An overview of VSA system design methodologies, including (a) an alternative hardware/software co-design framework; (b) a demonstration through a co-design case study.}
	\label{fig:VSA-design-methods}
\end{figure}

Despite advances in today's hardware developments, they still face limitations in adaptability to VSA models and their ability to generalize across different tasks and precision levels; such characteristics are inherent in today's multi-modal VSA-centric flows~\cite{hersche2023neuro}. Most current hardware lacks reconfiguration capabilities, thereby hindering systematic and scalable interplay between VSA algorithms and hardware. Typically, they employ a rigid, unstructured, and ad-hoc design methodology. This approach also cannot meet the growing demand for heterogeneous cognitive models, such as neuro-symbolic systems~\cite{wan2024towards,wan2024towards_2} and generative cognitive architectures~\cite{furlong2023bridging}. Specifically, it shows limited capability in integrating VSA dataflows with other computational primitives like tensor units~\cite{yang2023neuro}. These limitations prevent the full exploitation of HD computations for next-generation VSA systems. Bridging this gap requires a structured and unified design approach, leading to a systematic and efficient interplay between hardware and software, as shown in Figure~\ref{fig:VSA-design-methods}(a).

Realizing a unified design approach involves developing a flexible \emph{hardware-software co-design} framework that fully exploits the inherent qualities of VSAs: robustness to noise, element-wise independence, and versatility. To achieve this goal, this framework should support multiple VSA models or representations, enabling adaptability to diverse computational tasks~\cite{schlegel2022comparison}. Structured and unified integration, as illustrated in Figure~\ref{fig:VSA-design-methods}(b), requires the compact formulation of VSA kernels, consideration of algorithm-specific features and dataflows, and adherence to hardware constraints and performance limits. By streamlining interactions among these components, the framework allows VSA kernels to be effectively mapped to hardware fabrics. This process should be guided by thorough feature investigations to ensure optimal performance. Advanced design methodologies, such as modular and parameterizable hardware blocks, can facilitate this adaptability, and machine learning techniques can also be incorporated for real-time optimization and task-specific tuning to further enhance efficiency. Ultimately, this unified approach will create versatile and high-performance VSA systems capable of meeting the diverse demands of next-generation cognitive applications.

\subsection{Contributions of the Paper}
\label{subsec:contributions}

The purpose of this paper is twofold. First, it surveys key principles of VSAs, state-of-the-art hardware implementations, and algorithms, thereby complementing previous algorithm- or system-level reviews for VSAs~\cite{schlegel2022comparison, kleyko2022survey, kleyko2023survey, kleyko2022vector, chang2023recent}. Second, it provides a comprehensive and unified discourse on the convergence of hardware and algorithms, aiming to bridge the gap between theoretical explorations and the development of efficient VSA hardware architectures. We reason that an efficient realization of a broader set of VSA methods necessitates a holistic understanding of all system requirements~\cite{frenkel2023bottom}, ranging from algorithm-level features and kernels down to hardware-level characteristics and design opportunities (Figure~\ref{fig:VSA-design-methods}. The outcome of the proposed study is a holistic framework that lays out the principles of hardware-software co-design and integration in VSA systems. To the best of our knowledge, this is the \emph{first} paper to address co-design principles and attributes for VSA systems, aiming to inspire the design of next-generation cognitive computing systems.

\subsection{Organization of the Paper}
\label{subsec:organization}

The rest of this paper is organized as follows. We first review the principles of VSA computing and the different VSA methods (Section~\ref{sec:compute_model}), laying the foundation for subsequent hardware-software co-design. This is followed by a comprehensive discussion of hardware implementations of VSAs, analyzing analog, mixed-signal, and digital circuit design styles to date (Section~\ref{sec:hardware}). We then establish a co-design framework that facilitates synergistic interactions between algorithm and hardware layers of abstraction (Section~\ref{sec:codesign}). We further illustrate an application of this framework that aims to create embodied cognition in autonomous machines (Section~\ref{sec:cognition}). Finally, we examine research challenges and opportunities for future explorations (Section~\ref{sec:challenge}). 

\section{Principles of Vector-Symbolic Computing}
\label{sec:compute_model}

This section provides a comprehensive overview of vector-symbolic computing and its fundamental concepts. We delve into the definition and mathematical foundations of high-dimensional distributed representation (Section~\ref{sec:vsa-basics}). We also present key VSA learning paradigms, with examples illustrating the need for VSA algorithm-hardware co-design (Section~\ref{subsec:algorithms}).




\subsection{Computing with High-Dimensional Vectors}
\label{sec:vsa-basics}

\begin{figure}[!t]
	\centering
	\includegraphics[width=\columnwidth]{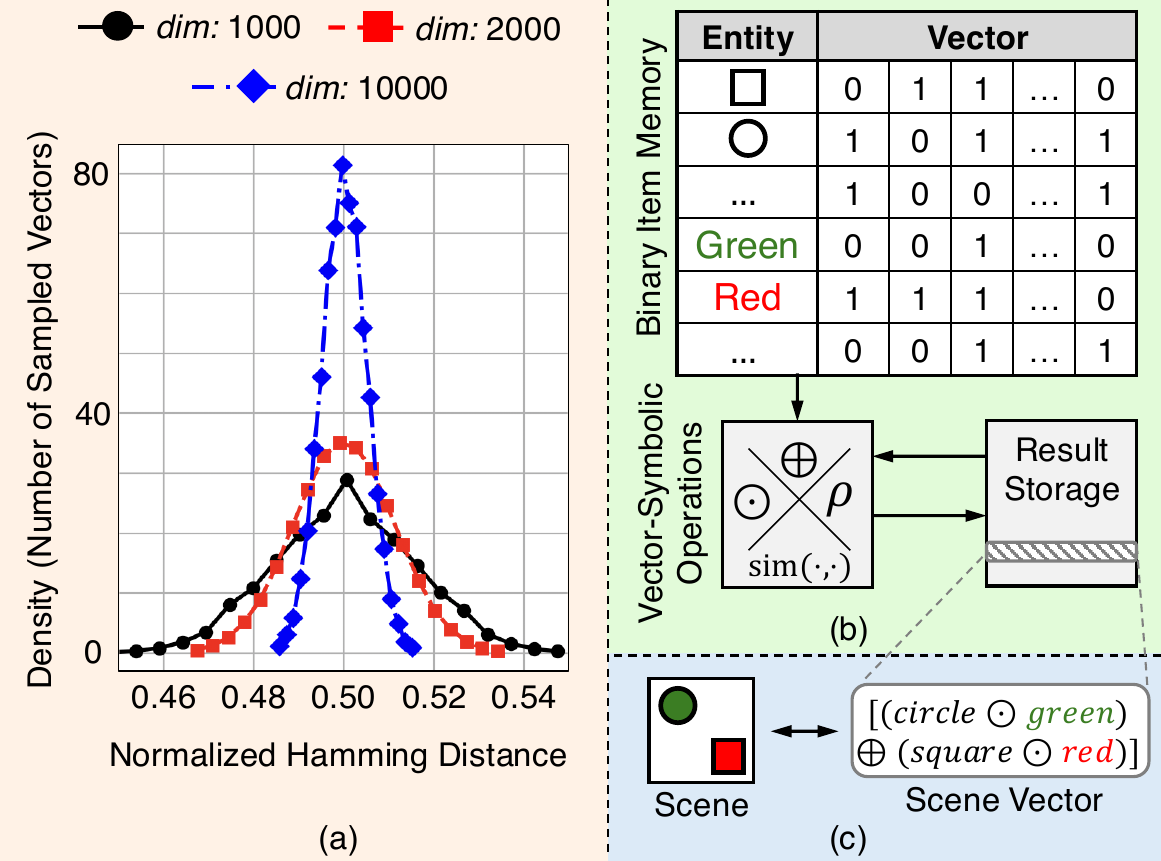}
	\caption{\textbf{Background of Vector-Symbolic Computing.} (a) Orthogonality in high dimensions becomes more pronounced, i.e., density concentration becomes sharper around 0.5, as vector dimensionality increases. (b) Basic elements of vector-symbolic computations: item memory and algebraic operations. (c) A superposed, higher-order scene vector constructed via vector-symbolic operations.}
	\label{fig:hdc_basics}
\end{figure}

Vector-symbolic computing is a method that encodes atomic attributes and patterns as high-dimensional distributed vector representations, commonly known as \textit{hypervectors}, which typically comprise 1,000 dimensions or more. The high dimensionality allows for a rich encoding space, enabling the representation of complex information with a high degree of redundancy, which contributes to robustness against noise. Hypervectors are generated using various methods, such as random initialization, where each element in the vector is randomly assigned a value from a predefined set or a probability distribution~\cite{kanerva2009hyperdimensional}. This initialization mechanism results in \textit{item hypervectors} that are nearly orthogonal to each other, as the likelihood of any two randomly generated hypervectors having a significant overlap is extremely low. The quasi-orthogonality feature becomes more pronounced as the length of the hypervectors is increased, as illustrated in Figure~\ref{fig:hdc_basics}(a). A collection of randomly generated item vectors is typically referred to as the \textit{item memory}.

Furthermore, vector-symbolic computing allows for constructing complex representations by combining hypervectors through basic mathematical operations, resulting in superposed vectors with the same dimensionality~\cite{gallant2013representing}. The basic mathematical operations used in such computations are vector similarity, binding, bundling, and permutation, as illustrated in Figure~\ref{fig:hdc_basics}(b). For instance, a visual scene can be represented by integrating hypervectors for individual objects and their attributes, as shown in Figure~\ref{fig:hdc_basics}(c). The resulting superposed vector retains the high-dimensional nature of the original hypervectors while embodying a richer and more nuanced representation of the combined elements. A brief description of vector-symbolic operations is presented below.

\noindent \textbf{Vector Similarity:} This operation, denoted by $\mathrm{sim}(\cdot, \cdot)$, quantifies how closely related two hypervectors are in the high-dimensional space. This measure can be based on various metrics, depending on the specific architecture and application. Common similarity measures include cosine similarity, Hamming distance, and dot product. Cosine similarity, computed as $\mathrm{sim}(a,b)=\frac{a\cdot b}{\|a\|\|b\|}$, ranges from $-1$ to $1$, indicating how aligned the vectors are. Alternatively, Hamming distance, which counts the number of differing bits between binary hypervectors, can be used for comparing vectors and is commonly adopted due to its amenability to low-cost hardware design~\cite{datta2023hdbinarycore, menon2022highly}. Similarity measures help determine the degree of resemblance between vectors and are essential for all vector-symbolic applications.


\noindent \textbf{Binding:} Binding, denoted by $\odot$, is an operation designed to model how the brain \textit{connects} input information as key-value pairs. This operation operates on two input hypervectors and generates another hypervector that belongs to the same space (e.g., $x=a \odot b$). An example implementation of this operation is XOR, which is often used when binding binary hypervectors. The generated output hypervector $x$ is quasi-orthogonal to its inputs ($a$ and $b$). This property is attributed to the fact that the binding operation does not preserve intrinsic vector representations; rather, it can only preserve \textit{similarity information} that governs vector relationships~\cite{rachkovskij2022recursive}: $\mathrm{sim}(a \odot c, b \odot c) \approx \mathrm{sim}(a, b)$. 

On the other hand, the inverse operation for binding, also called the release or unbinding operation ($\circleddash$), can retrieve information by disentangling the joint representation~\cite{kleyko2023efficient}. For instance, if we have a vector $x$ obtained by binding $a$ and $b$ ($x=a\odot b$), applying the unbinding operation $x \circleddash b$ will produce a vector that closely approximates $a$. Specifically, $\mathrm{sim}(x \circleddash b, a) \gg \mathrm{sim}(x \circleddash b, \epsilon)$, where $\epsilon$ is a random hypervector. This formula indicates that the unbinding operation effectively recovers the original vector $a$ or a close approximation from the composite vector $x$, with high similarity. 

\noindent \textbf{Bundling:} Bundling, denoted by $\oplus$, is an operation designed to model how the brain \emph{memorizes} input information as a set of entities. This operation fuses information from all input hypervectors and generates a hypervector that represents their mean, that is, the output is maximally similar to all of the inputs: $\mathrm{sim}(a\oplus b\oplus c, a) \gg \mathrm{sim}(a\oplus b\oplus c, \epsilon)$, where $\epsilon$ is a random hypervector. The bundling of binary hypervectors is typically implemented using a simple addition operation and then applying a majority function over the sum vector to generate a binary vector. In some implementations, the summed vector is not binarized, preserving its higher-resolution structure~\cite{teeters2023separating}. The downside of such implementations is the significant increase in computational and memory requirements, as the representation must store and process real-valued or high-bit-depth vectors rather than binary ones.

\noindent \textbf{Permutation:} The permutation, denoted by $\rho(\cdot)$, is a unary operation that involves deterministically reordering the elements of a given hypervector, such as through cyclic rotation of the whole vector. This operation is essential for encoding sequences and ordered data, allowing hypervectors to represent the temporal or spatial arrangement of features and events~\cite{joshi2017language, karunaratne2021energy}. For example, let $x_g$, $x_o$, and $x_d$ be the hypervector representations of the letters ``g'', ``o'', and ''d'', respectively. To encode the sequence of letters in the word ``good'', permutation is applied to each subsequent letter's hypervector and then combined. Specifically, the word ``good'' can be constructed as $x_g\odot\rho(x_o)\odot\rho^2(x_o)\odot\rho^3(x_d)$, where $\rho^k(\cdot)$ denotes the $k$-th permutation applied to the hypervector. Note that an arbitrary vector $x$ and its permutation $\rho(x)$ are quasi-orthogonal to each other, meaning they can be easily distinguished from one another in the high-dimensional space. 

\subsection{VSA Algorithm Domains}
\label{subsec:algorithms}

\noindent \textbf{Supervised Learning:}
Supervised learning with VSA has been proposed during the early stage of the development of its framework~\cite{kanerva1988sparse}, which is also the foundation of subsequent VSA-inspired applications. Steps in VSA classification can be categorized into three steps: \textit{encoding}, \textit{training}, and \textit{inference}. In the encoding phase, the feature $\vec{F}_{1\times n}$ is first projected to high-dimensional vector (hypervector) $\vec{H}_{1\times N}$ by multiplying the encoder $\vec{E}_{n\times N}$, with the element in the encoder sampled with random distribution. These encoders create an embedding $\phi: \mathbb{R}^n\rightarrow\mathbb{R}^N$ such that $<\phi(x), \phi(y)> \approx k(x,y)$, where k is a shift-variant kernel~\cite{thomas2021theoretical}. Most of the existing VSA encoders originate from the fact that the Fourier transform of a shift-invariant kernel k is a probability measure, a result that originates from Bochner's Theorem~\cite{rudin2017fourier}.

After the encoding phase, training is performed to form the class hypervector $\vec{C}$. VSA is designed as a hardware-friendly framework that supports \textit{single-pass} training. In the single-pass training phase, all the hypervectors belonging to label $l$ are aggregated to form $\vec{C_l}$. To further increase the accuracy of the VSA framework, inspired by the perceptron learning algorithm, the \textit{iterative} training scheme subtracts the wrongly predicted class hypervector $\vec{C_{l'}}$ and adds the correctly predicted class hypervector $\vec{C_{l}}$:
\begin{equation}
\begin{aligned}
    C_l &\leftarrow C_l + \eta (1-\delta)\\
    C_{l'}&\leftarrow C_{l'} - \eta (1-\delta)
\end{aligned}
\end{equation}
where $\eta$ is the learning rate for fine-tuning and $\delta$ is the similarity value (e.g., cosine similarity) of the class hypervector and the query. After the training, the VSA model is ready for inference, where the query feature is first processed with the same encoder for projecting into high-dimensional space, and then searched across all the class hypervectors with a pre-defined similarity metric as mentioned in Section~\ref{sec:vsa-basics}. Finally, the highest similarity value is the predicted class for this query. Existing VSA hardware designs focus mostly on accelerating similarity search, as such an operation dominates the inference run-time. However, there are cases that the encoder dominates the inference run-time, such as time-series encoding~\cite{ni2022neurally}, illustrating the need for \textit{top-down perspectives} on the VSA co-design framework.

\noindent \textbf{Unsupervised Learning:}
Unsupervised learning with VSA involves investigating the inherent statistical relationship between data in their high-dimensional representations. Imani et al. proposed HDCluster that adopts the K-means principle to generate clusters through iterations of comparison and refinement~\cite{imani2019hdcluster}. Central hypervectors are randomized evenly within HD space and iteratively updated by averaging all hypervectors annotating the same label. Learning phase ceases when the step of update becomes negligible. Computational efficiency could be improved through weighted updating of central hypervectors with the confidence level generated through cluster assignment phase~\cite{hernandez2021framework}. Much resembling Deep Neural Networks, training and clustering (similarity search) could be performed with various types of operand: floating-point central hypervectors permitted more accurate updates ( $C_{l} \leftarrow C_{l}+x$ ), while binarized hypervectors enabled more hardware-friendly similarity search using hamming distance ($sim(C^{b}_l , x)$). Applications in other realms include spectrum clustering~\cite{xu2023hyperspec} and traffic profiling~\cite{bandaragoda2019trajectory} which demonstrate the efficiency and efficacy of VSA-based unsupervised learning.

On top of Holographic Reduced Representation (HRR)~\cite{plate1994distributed}, hyperseed ~\cite{osipov2022hyperseed} facilitates unsupervised learning through optimizing a projection hypervector $s$ between the encoded input hypervector space $D$ and a mapped hypervector space $P$ in complex domain ($D \circleddash s \rightarrow P$). In each training iteration, the most divergent input hypervector is identified and assigned an anchor in the mapped hyperplane, representing a center for similar inputs. Such mapping informatics is recorded by incrementing projectory hypervector with the anchor-data binding pair $s \leftarrow d \odot p + s$. Since (un)binding preserves similarity between hypervectors, the mapping capability would be enhanced to project input hypervectors similar to $d$ to the newly designated center. Through incorporating straying outliers, the mapped hyperspace is gradually distributed with clusters. Hyperseed achieved comparable performance against self-organizing map (SOM)~\cite{chaudhary2014novel} while significantly reducing the number of running passes by alleviating parallel search overhead with neuromorphic hardware~\cite{liu2022cosime}.




\section{VSA Hardware: Architecture and Technology}
\label{sec:hardware}
To enhance the efficiency of VSA-based applications with optimized HD compute and memory kernels, leveraging tailored architectures along with appropriate device technologies is crucial. This section reviews hardware implementations for VSAs from digital platforms (e.g., microprocessors, FPGAs, etc.) to custom ASIC designs exploiting in-memory or near-memory computing architectures and emerging device technologies. The mapping between these hardware designs and the kernel functionalities is discussed, along with the corresponding design tradeoffs.

\subsection{Digital Hardware Platforms}
With high overheads and a lack of VSA-specific parallelism, floating-point arithmetics in CPUs and GPUs do not match efficiently with the high-dimensional vector algebra for VSA applications, implying the need for HD-vector-centric processor architectures~\cite{datta2019programmable}. Taking advantage of the fine-grained parallelism and pipelining capabilities of FPGAs~\cite{liu2021robotic,liu2022energy}, \cite{salamat2019f5} introduces an automated FPGA framework for HD computing acceleration. \cite{menon2023accelerating} relies on a novel bit-serial word-parallel approach to enhance the spatial encoder in VSA bundling. \cite{montagna2018pulp} proposes HD computing acceleration on the Parallel Ultra-Low Power (PULP) Platform, a software programmable cluster architecture, with a parallel processing chain method that separately parallelizes each component in the chain and distributes the computational tasks across multiple processing cores for performance optimization.  

In addition to embedded system prototypes, custom ASICs have also been developed for hardware acceleration of HD computing. \cite{datta2019programmable} proposes a programmable HD processor design utilizing a Hyper-dimensional Logic Unit (HLU) systolic array architecture for the HD encoder. This leads to a reconfigurable architecture capable of forming multiple HLU layers, and exhibits excellent energy efficiency across various supervised classification tasks, including language recognition and human face detection. In~\cite{datta2023hdbinarycore}, the HDBinaryCore, a digital 28 nm CMOS chip, is the first silicon prototype as a programmable HD processor for biosignal processing. It adopts the architecture proposed in~\cite{datta2019programmable}, allowing programmability through the specification of interconnection and operation for each HLU Layer.

\subsection{In-Memory and Near-Memory Computing Architectures}

The high-dimensional nature of hypervectors in VSA operations can incur intensive and frequent data movement between processing units and memory components, creating a bottleneck for VSA applications within traditional von Neumann architectures. Meanwhile, the similarity search among high-dimensional vectors as a key building block in VSAs inspires memory-centric designs. Benefiting from highly efficient in-/near-memory operations and significant computational parallelism, emerging in-memory computing (IMC) or near-memory computing (NMC) architectures present a promising path to address the VSA-specific memory walls. These architectures, specifically designed for VSAs, fall into two primary categories: digital IMC/NMC~\cite{wang2023computing,gupta2018felix,dutta2022hdnn} and analog IMC/NMC designs~\cite{karunaratne2020memory,morris2021hydrea,morris2022stochastic,hsu2022memory}.

Digital IMC architectures typically employ binary in-memory or near-memory logic circuits to accelerate VSA operations. Thus, the digital nature of these computations can eliminate the accuracy loss induced by analog-to-digital conversions. \cite{wang2023computing} proposes an HD outlier detection (ODHD) accelerator combining SRAM IMC processing element (PE) array with assistant NMC, which enables high throughput VSA operations (binding, bundling, permutation) in the ODHD algorithm due to parallel computation across the different subarrays. Specifically, the digital SRAM PEs are co-designed with the ODHD algorithm for efficient mapping of encoding, training, threshold calculation, and fine-tuning steps. \cite{gupta2018felix} presents an HD computing accelerator with a resistive RAM (RRAM) IMC crossbar capable of single-cycle NOR, NOT, NAND, and OR logical functions in memory. This RRAM array efficiently stores pre-trained identity hypervectors and quantized hypervectors within a single memory partition, performing in-memory XOR operations in series. This process maps a feature vector with n elements to high-dimensional space, enabling acceleration in the HD encoding module. In~\cite{dutta2022hdnn}, the HDnn-PIM architecture is introduced as a high-dimensional neural network (HDnn) IMC solution. Specifically designed for complex image classifications, it incorporates a few initial stages of convolution-based feature extraction to enhance the effectiveness of HD learning on intricate data. The HDnn-PIM employs a tiled architecture with RRAM crossbars, featuring supertiles that facilitate input reuse and fusion of tiles for handling large input sizes.

Architectures for analog in/near-memory computing leverage physical principles such as Ohm's law or charge sharing/re-distribution. These principles are utilized for two fundamental operations: multiplication and addition. \cite{karunaratne2020memory} proposes an in-memory HD computing system implemented on two phase-change memory (PCM) crossbar engines alongside peripheral digital CMOS circuits. The first PCM crossbar engine is dedicated to accelerating HD encoding, employing in-memory read logic operations for hypervector binding and near-memory CMOS logic for hypervector bundling. The second PCM crossbar engine is employed for associative memory (AM) search, utilizing in-memory dot-product acceleration to calculate the inverse Hamming distance. An experiment involving 760,000 PCM devices for analog IMC also demonstrates comparable accuracies to software implementations. In~\cite{morris2021hydrea}, an analog RRAM IMC architecture is proposed, capable of supporting dot-product for matrix multiplication in encoding and similarity search during inference. Additionally, it supports addition and subtraction during training and retraining to enhance model robustness. As an example of algorithm-hardware co-designs, ~\cite{morris2022stochastic} introduces a stochastic HD computing system, leveraging the complex task-solving capabilities of VSA models.

A growing trend is that emerging non-volatile memories (NVMs) are being widely explored and adopted in IMC designs for VSA systems~\cite{zheng2024cmos+}, including RRAM~\cite {morris2022stochastic,dutta2022hdnn,gupta2018felix,morris2021hydrea,crafton2022improving}, PCM~\cite{karunaratne2020memory}, and 3D NAND Flash~\cite{hsu2022memory}. The on-chip NVMs provide fast, energy-efficient access and compute, whereas the off-chip 3D Flash offers even higher storage capacity. Non-volatility benefits edge VSA systems for lifelong learning and inference~\cite{li2021sapiens}. NVMs can serve as associative memories in encoding and similarity search operations, which only require infrequent memory writes. The biggest challenge with NVMs in conventional IMC designs targeting DNNs is that the device non-idealities may heavily impact model accuracy. In contrast, given the high-dimensional and holographic nature of VSAs, error bits in hypervectors are not contagious, leading to a computational system inherently resistant to defects and disturbances. In~\cite{li2016hyperdimensional}, multi-layer 3D vertical RRAM (VRRAM) integrated with FinFETs forms energy-area-efficient MAP kernels, exploiting vertical connectivity of heterogeneous device technologies. In~\cite{wu2018hyperdimensional}, a highly-efficient, end-to-end 3D HD nanosystem is built, leveraging the integration of RRAMs and carbon nanotube field-effect transistors (CNFETs). The inherent variations in RRAMs and CNFETs are collectively harnessed while demonstrating robustness to errors inherent in the underlying hardware.



\subsection{System Integration and Scaling}
Scaling up memory-centric designs in a cost-effective fashion may further make VSA systems more efficient by providing higher total capacity and exposing additional parallelism. Technology scaling and integration of emerging memories will continue to unlock additional system optimization opportunities. Rapid industry progress has been observed, for example, the back-end-of-line (BEOL) integration of state-of-the-art RRAM technologies in foundry FinFET processes~\cite{golonzka2019non, lee2020memory}. Blurring the on-chip vs. off-chip boundary and leveraging the system integration techniques will help amortize the costs of mapping large-scale VSA workloads or processing large databases. 
Advanced packaging and heterogeneous integration will become increasingly important in enabling flexible designs where VSA hardware fabrics may leverage diverse semiconductor technologies to improve system-level connectivity, energy efficiency, and footprints. Traditional 2.5D and 3D integration involves stacking and bonding separate chips, utilizing through-silicon vias (TSVs) and micro-bumps for vertical connections. Typically, TSVs have a pitch ranging from 20 to 40 $\mu$m, with advanced designs achieving pitches as fine as 10 $\mu$m, and the pitch for micro-bumps is approximately 40 $\mu$m~\cite{shulaker2017three}. Further scaling down the micro-bump pitch can be challenging for higher interconnection bandwidth. However, hybrid bonding technology~\cite{agarwal20223d} is being actively developed and can be integrated with backend wafer/die packaging, enabling higher interconnection density, energy efficiency, and improved signal/power integrity. 

Heterogeneous 3D integration can provide unique functionalities and performance enhancement for VSA systems. For instance, 2.5D and 3D designs with heterogeneous chiplets embedded with computational memories could offer significant improvements in energy efficiency and energy-delay product (EDP) compared to conventional 2D architectures ~\cite{du20253dcimlet}. Furthermore, the on-sensor computation~\cite{liu2022augmented} is facilitated by the triple-wafer-stacking technology, where a logic wafer is integrated with dual-layer stacked digital pixel sensors through micro TSVs, and sensor layers are face-to-back hybrid bonded. The integration of sensors and logic dies can further facilitate VSA-based robotic applications, interweaving perception, reasoning, and control modules. In addition, for large-scale VSA architectures that need high data bandwidth and memory capacities, 3D DRAM (e.g., high bandwidth memory (HBM) utilizing DRAM die stacking) may play a crucial role in co-designing future VSA computing systems. 

As a complement to heterogeneous integration, monolithic 3D integration interweaves layers of sensors, memories, and logic through monolithic inter-layer vias (MIVs). The diameters of MIVs are much smaller compared to TSVs~\cite{gopireddy2019designing}, and the pitch of MIV can reach 100 nm in 14 nm process~\cite{samal2016monolithic}. Therefore, monolithic 3D integration enables close integration between logic and memory components and facilitates much denser vertical connections than TSVs~\cite{10966007,10873546}. In~\cite{wu2018hyperdimensional}, CNFETs and RRAM are deployed for logic circuits and memories, respectively, and are integrated using the monolithic 3D technology, which is feasible due to the low-temperature fabrication of RRAM and CNFET.

VSA system integration may further benefit from alternative chip-to-chip communication. For instance, wireless in-package communication technology was used to interconnect a large number of physically distributed IMC cores, allowing for joint broadcast distribution and computation, and extensive parallelization of the architecture to improve system performance~\cite{guirado2022wireless,guirado2023whype}.
\section{VSA Hardware/Software Co-Design}
\label{sec:codesign}

To cope with the complexity of next-generation VSA systems, we reason that an effective design process should comprise two essential steps: (1) \emph{kernel formulation}, which aims to formally\footnote{The formulation may involve mathematical descriptions, flow charts, or a representation using data structures.} describe the dynamics of the target function and map this formulation into abstract data and control flows; (2) \emph{feature-aware hardware realization}, which takes the abstract data and control representations along with cross-layer features and map them into suitable hardware components (Figure~\ref{fig:VSA-design-methods}(a)). It is worth noting that the first step provides a technology-agnostic procedure, whereas the second step focuses on incorporating the specifics of the problem domain. Below, we discuss these two steps in detail and present a unified co-design framework, bringing these two steps into a merged design space for VSA-based cognitive systems.

\subsection{VSA Kernels}
\label{subsec:kernels}

The first step in the design process is to determine the VSA kernel functions that need to be realized in hardware. A major advantage of kernel formulation is its ability to efficiently capture multiple sequences of high-dimensional operations and also present proper control primitives for selecting configurations. The goal here is to represent the various dynamics and interactions between hardware and software while abstracting away low-level details of the hardware~\cite{ibrahim2024efficient}. This is typically achieved by formulating the kernel computation as $F(I,P,C)$, where $F(\cdot)$ represents a collective array of all target kernel functions $\{f_1, \ldots, f_m\}$, which together cover the whole domain of high-dimensional operation sequences. The argument $I$ is an array of randomly generated item hypervectors, and $P$ is an array of composed hypervectors. 
The argument $C$ represents a set of conditional constructs that define the subdomains associated with the computational elements $f_i$. As such, when mapping the formulation to hardware, the set $C$ conceptualizes multiplexers and other control primitives employed to select HD data paths according to the execution mode. This methodology offers a high degree of flexibility and is commonly used with domain-specific processors~\cite{sze2017efficient, srivastava2018promise}. 

\begin{figure}[!t]
	\centering
	\includegraphics[width=0.9\columnwidth]{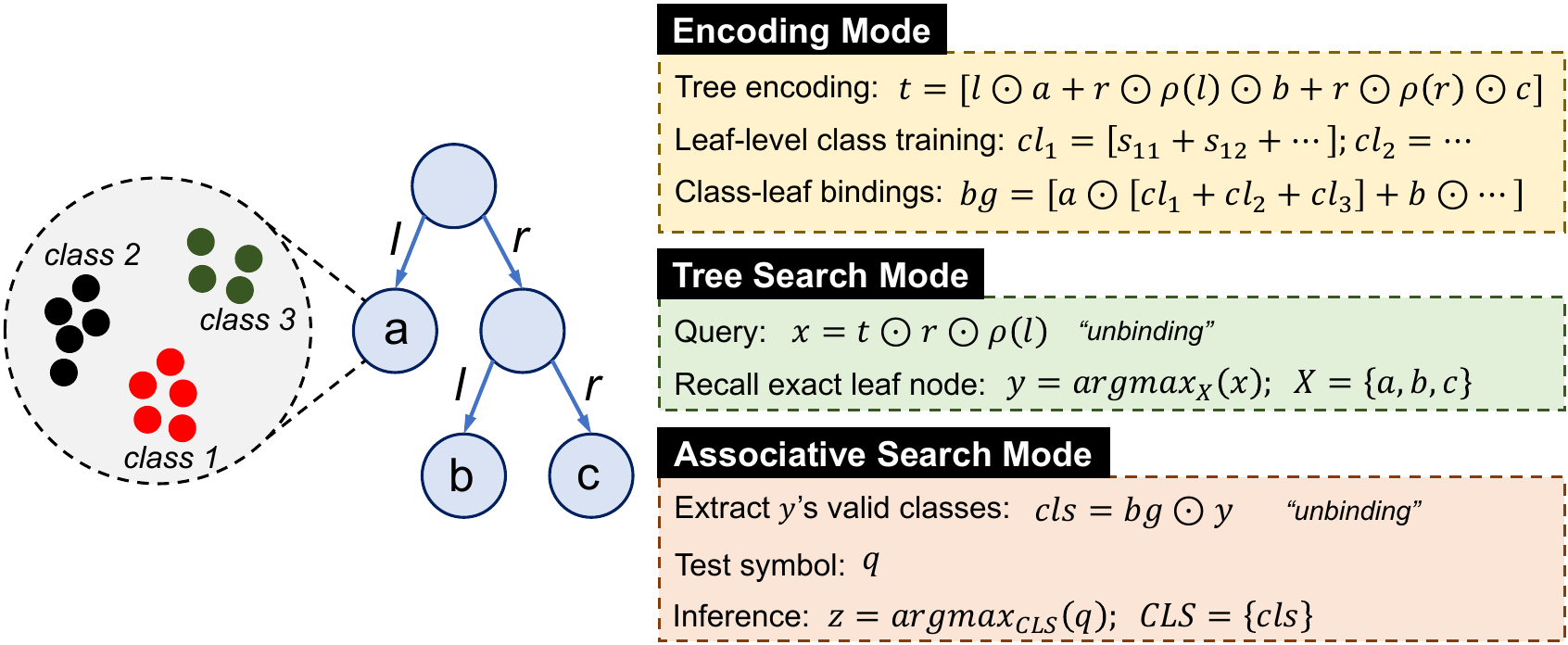}
	\caption{An example of an arbitrary VSA classification application with its three operating modes: encoding, tree search, and associative search.}
	\label{fig:kernels-ex}
\end{figure}

\noindent \textbf{Example:} To illustrate kernel formulation, we consider an arbitrary VSA classification algorithm that performs pattern recognition, which is distributed across a tree structure. Particularly, as illustrated in Figure~\ref{fig:kernels-ex}, the algorithm has three modes: \emph{encoding}, \emph{tree search}, and \emph{associative search}. The encoding mode aims to construct a tree of classes, where each leaf node represents a local classification setting. As such, each leaf node receives symbols and encodes them into a unique set of classes. In the tree search mode, the goal is to determine the identity of a leaf node given the path leading to it from the root. The associative search operates at the leaf level and aims to produce a class label given a test symbol. 

\begin{figure}[!t]
	\centering
	\includegraphics[width=0.9\columnwidth]{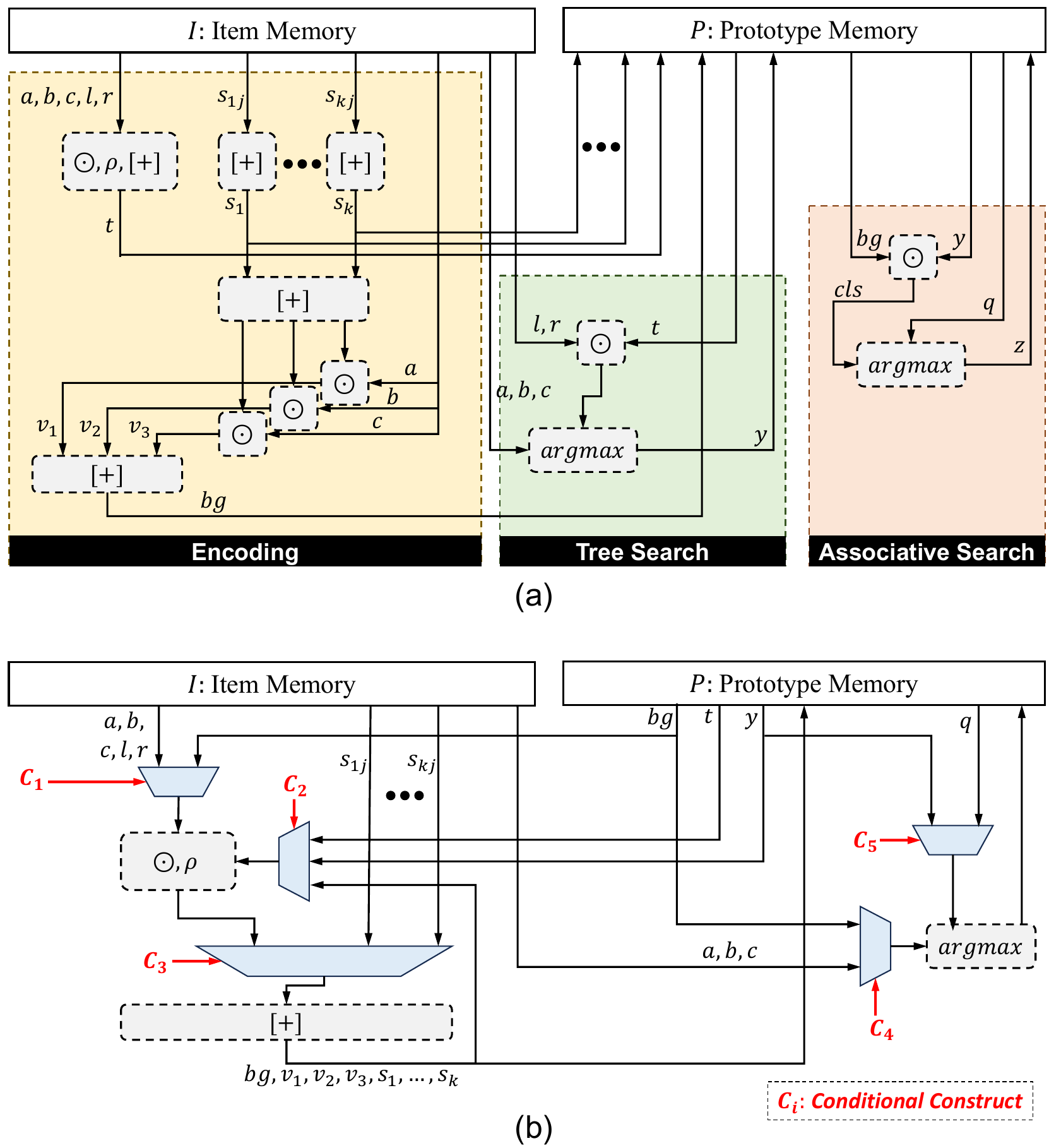}
	\caption{Visualization of kernel formulations for the example in Figure~\ref{fig:kernels-ex}: (a) spatial kernel formulation; (b) temporal kernel formulation.}
	\label{fig:kernel_chart}
\end{figure}

Given the characteristics described above, we realize that multiple kernel formulations $F(I,P,C)$ can be inferred for such an application. For example, Figure~\ref{fig:kernel_chart}(a) shows a spatial kernel formulation where computations for all the individual modes are distributed spatially. This approach does not require any control conditions in order to set up the data paths for execution, i.e., the set $C$ is empty. In other words, this approach of kernel formulation results in a highly parallel hardware design with very limited software reconfigurability. 

On the other hand, Figure~\ref{fig:kernel_chart}(b) shows a temporal kernel formulation where computations for all the individual modes are centralized, resulting in data paths that are shared among all execution modes. This approach requires a comprehensive set of control constructs $C$ to be incorporated into the data paths. As such, this approach results in an area-efficient design with a high degree of software reconfigurability. However, the resource sharing adopted by this approach may result in a significant increase in latency compared to the approach in Figure~\ref{fig:kernel_chart}(a). 

It is evident that the choice of a kernel formulation can significantly impact the overall performance and efficiency of the application. The selection of a specific kernel formulation determines how computational resources are utilized, influencing aspects such as parallelism, data flow, and hardware-software interaction. Moreover, the adaptability and reconfigurability of the chosen kernel formulation play a crucial role in addressing dynamic requirements and optimizing resource utilization. A well-chosen kernel formulation aligns with the specific characteristics and demands of the application, striking a balance between hardware efficiency and flexibility. Similar tradeoffs can also be found when choosing hardware technologies to implement a kernel formulation---some details on this matter will be presented later in Section~\ref{subsec:hardware_realization}.

\subsection{Cross-Layer Features and Hardware Realization} 
\label{subsec:hardware_realization}

Besides kernel formulation, the design process also requires an exploration of essential \emph{domain-specific properties}---a set of key attributes that are primarily inspired by the dynamics of VSA methods. Exploring such properties plays a key role in bridging the gap between theoretical aspects of VSA methods and their counterparts in hardware features. Below, we elaborate on each of the domain-specific properties, with the associated cross-layer features summarized in Table~\ref{table:codesign_attributes}. 

\begin{enumerate}
\item \emph{Leveraging Stochastic Characteristics:} VSA methods are not inherently classified as stochastic computing architectures; nevertheless, they exhibit stochastic-like characteristics that, when harnessed, can lead to enhanced computational efficiency~\cite{poduval2021stochd,wan2024h3dfact}. An illustrative instance is the utilization of controlled noise-injection techniques, which prove beneficial in breaking limit cycles and enhancing the efficiency of recurrent VSA algorithms, such as resonator networks~\cite{frady2020resonator}. Such stochastic techniques are only achievable through synergistic hardware/software integration. In other words, there is a need to align the probabilistic essence of VSAs with hardware primitives that can inject controlled noise, e.g., via ring oscillators, or are inherently stochastic~\cite{langenegger2023memory}. 

\item \emph{Controlling Performance-Accuracy Tradeoffs:} VSAs offer a set of intrinsic control knobs that can finely adjust the system's performance and accuracy behaviors in alignment with application requirements. Key control knobs include vector length, sparsity level, binding complexity, and thresholding~\cite{kleyko2018classification}. Taking vector length as an illustrative example, a shorter vector length enables faster and more energy-efficient processing, rendering it well-suited for real-time applications or scenarios with stringent latency and energy constraints. Conversely, a longer vector leads to enhanced superposition capacity, thus potentially improving the accuracy~\cite{kleyko2022vector}. 
However, this advantage comes with increased computational complexity and hardware area. Examples of such techniques are spatial tiling~\cite{datta2019programmable} and temporal vector folding~\cite{menon2021highly}. As such, identifying insignificant dimensionality in hyper-vectors has been proposed recently in~\cite{zou2021scalable}. With the observation that in VSAs, not all dimensions would have similar impact on the learning task. There are some dimensions with minimal impact during the learning phase, underscoring the possibility of maintaining both hardware performance and software accuracy. Other examples that could maintain both hardware performance and software accuracy include novel quantization scheme~\cite{barkam2024in} and sparsity-aware hardware design~\cite{chen2023hypergraf}.


\item \emph{Adaptability to Multiple Cognitive Tasks:} As detailed in Section~\ref{sec:compute_model}, the mathematical properties inherent in VSAs impart them with the versatility to tackle a wide array of tasks, encompassing optimization, factoring, classification, as well as learning and reasoning. However, seamless integration of various cognitive tasks demands an interplay between flexible software implementations and adaptable hardware dataflows. On the software front, achieving flexibility involves constructing hierarchical cognitive architectures within VSAs, facilitating the organization and processing of encoded information across different levels of abstraction~\cite{olascoaga2022brain}. Meanwhile, from a hardware perspective, the efficient sharing of memory-search components and high-dimensional arithmetic units becomes crucial, tailored to the specific demands of diverse cognitive tasks~\cite{ibrahim2024efficient}. A comprehensive co-design framework for VSAs should consider the joint-tuning of both software flexibility and hardware reconfigurability to exploit the diversity of VSA methods.

\item \emph{Interoperability with Real-World Interfaces:} The integration of cognitive systems with real-world environments relies heavily on their ability to navigate diverse multi-modal sensory-motor interfaces~\cite{sensinger2020review}. The inherent diversity of these interfaces leads to heterogeneous representations, often entangled within a low-dimensional space. The adoption of VSAs in this context allows these representations to be effectively mapped to high-dimensional spaces, thereby achieving linear separability. This transformation, often referred to as \emph{encoding}, leads to an optimized realization of sensory-motor applications, especially since only a few samples need to be processed through VSA operations~\cite{mitrokhin2019learning, schindler2021primer}; refer to Section~\ref{subsec:algorithms}. Yet achieving effective VSA encoding necessitates seamless coordination between hardware knobs---specifically, sampling frequency and dataflow capacity---and other software knobs, such as the encoding complexity and supported quantization levels~\cite{thomas2021theoretical, menon2021highly}.

\item \emph{Compatibility with Neural Cognitive Systems:} VSAs exhibit a compositional nature through their ability to represent complex symbols by combining simpler symbols in a structured and hierarchical manner. This feature renders VSAs highly compatible with neural network representations, giving rise to a class of models referred to as \emph{neuro-vector-symbolic architectures} (NVSAs)~\cite{hersche2023neuro,zhang2021abstract}. These integrated models enable sophisticated and generalized probabilistic reasoning, thus surpassing the limitations of neural networks that operate in isolation~\cite{Menon2022, karunaratne2021robust}. However, the true benefit of realizing NVSA models emerges only when their inherent compositionality is mirrored in the hardware realizations. Specifically, it is necessary to ensure that computational units and memories originally designed for accelerating neural networks can also be shared, or at the very least accessed, by dataflows that perform vector-symbolic operations. This synergy between software and hardware is key for advancing NVSAs~\cite{yang2023neuro}.
\end{enumerate}

\begin{table}[!t]
\caption{Representative cross-layer features that address VSA's domain-specific properties.}
\small
\centering
\resizebox{\linewidth}{!}{%
\renewcommand{\arraystretch}{1.3}
\begin{tabular}{c|l|l|l}
\hline
    \multicolumn{1}{c|}{\textbf{Properties}} & 
    \multicolumn{1}{c|}{\textbf{Software Features}} & 
    \multicolumn{1}{c|}{\textbf{Hardware Features}} & 
    \multicolumn{1}{c}{\textbf{Use Cases}}  \\
\hline
    \multirow{2}{*}{{[1]}} & 
    Random bit-flipping (noise) & 
    Inherent variations in NVMs & 
    Resonator networks~\cite{langenegger2023memory} \\ 

    &
    Stochastic arithmetic &
    Logic-in-memory & 
    Feature extraction~\cite{poduval2021stochd}\\

    
\hline    
    \multirow{5}{*}{{[2]}} & 
    Variable vector lengths & 
    Temporal vector folding & 
    Vector generation~\cite{menon2021highly}\\ 

    &
    Dropping insignificant dimensions &
    Noise-energy co-optimization&
    IoT system~\cite{zou2021scalable}\\

    &
    Controlled sparsity level& 
    Density-thinning logic& 
    Sparse event recognition~\cite{hersche2020integrating} \\ 

    &
    Scheduler for sparse graph&
    Pipeline-style decoder IP&
    Graph reasoning~\cite{chen2023hypergraf} \\
    &
    Non-linear quantization &
    Learning realistic device noise&
    Genome sequence matching~\cite{barkam2024in}\\
\hline 
    \multirow{1}{*}{{[3]}}& 
    Hierarchical models & 
    Programmable dataflows& 
    Sensory-motor learning~\cite{ibrahim2024efficient}\\  
\hline 
    \multirow{2}{*}{{[4]}} & 
    Variable quantization levels & 
    Data-level parallelism & 
    Online perception~\cite{mitrokhin2019learning} \\ 

    &
    Stream data processing&
    On-chip encoding& 
    Seizures detection~\cite{pale2022multi}\\
\hline 
    [5]& 
    Neuro-vector integration & 
    Hybrid vector/scalar dataflows &
    Visual reasoning~\cite{hersche2023neuro}\\

\hline
\end{tabular}
}
\label{table:codesign_attributes}
\end{table}

\begin{figure}[!t]
	\centering
	\includegraphics[clip, trim=0.1cm 0.1cm 5cm 0.1cm, width=1.00\columnwidth]{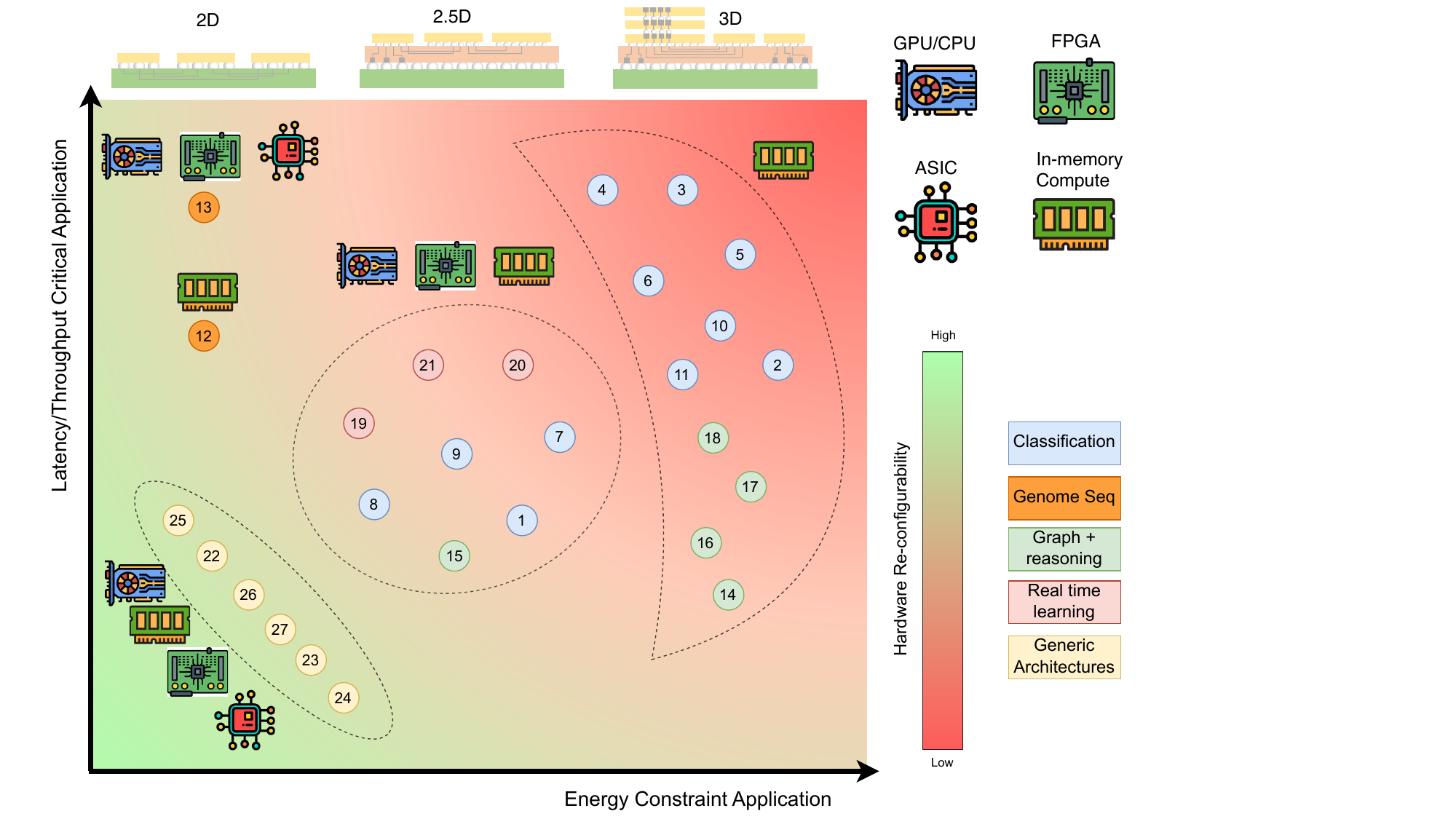}
        \caption{Mapping of VSA compute platforms to application domains based on latency/throughput and energy constraints.}
	\label{fig:hardware-mappings}
\end{figure}

It is worth noting that our goal here is not to identify a comprehensive set of hardware features; instead, we aim to outline guidelines or a methodology for deriving such features. As such, we seek to provide a representative set that can be incorporated into the co-design process based on the complexity or performance of the target context, as illustrated in Figure~\ref{fig:hardware-mappings}. We map possible implementations in a two-dimensional space, categorizing them by their latency/throughput and energy consumption requirements across various applications. Each data point, derived from related research as referenced in Table~\ref{table:hardware_map}, is color-coded to match its corresponding application area.

\begin{enumerate}
\item \emph{Classification Applications:}
Classification stands out as one of the most compelling applications of VSA computing, with various applications demonstrating a critical sensitivity to factors such as latency and energy consumption. Examples of VSA classification benchmarks include physical activity classification, speech recognition, cardiotocogram classification, and applications targeted towards human-centric Internet of Things (IoT) systems~\cite{datta2019programmable}. The encoder and associative search modules are pivotal in the classification process. Recent research efforts have embraced in-memory computing, employing both digital~\cite{wang2023computing, datta2019programmable} and analog/mixed-signal~\cite{wu2018hyperdimensional, morris2022stochastic} approaches, to enhance the efficiency of the encoding process and associative memory searches. These innovative designs, however, may encounter limitations in reconfigurability. For instance, some encoders are tailored to specific data types~\cite{dutta2022hdnn} or encoding patterns~\cite{datta2019programmable}, while others exploit emerging device technologies such as PCM~\cite{karunaratne2020memory} and 3D integration of CNFET and RRAM~\cite{wu2018hyperdimensional}. Such advancements often require the development of application-specific custom solutions to fully realize their potential. 

\item \emph{Genome Sequencing Applications:}
The large-scale databases required for genome sequencing applications present significant throughput and capacity challenges for hardware. To tackle these issues, researchers have employed high-density 3D NAND Flash to efficiently handle the large datasets of class hypervectors~\cite{hsu2022memory}. In addition, the integration of reference buffers with parallel computing units has proven effective in alleviating memory access bottlenecks, reducing reliance on large on-chip caches, and enhancing overall system performance~\cite{kim2020geniehd}. 

\item \emph{Graph and Reasoning Applications:}
Graphs naturally represent relationships and interactions across domains such as social networks, biological systems, and knowledge graphs. Reasoning over these graphs encompasses tasks like graph reconstruction, node classification, and graph matching. Addressing these tasks has led to the development of domain-specific encoder and decoder hardware~\cite{chen2023hypergraf}, with a particular emphasis on exploiting data sparsity to improve computational efficiency~\cite{kang2022relhd, chen2023hypergraf}. Additionally, emerging hardware technologies like FeFET and PCM are being explored~\cite{shou2023see}. 
The inherent device noises~\cite{barkam2023reliable}, stochastic properties~\cite{langenegger2023memory}, and switching dynamics~\cite{poduval2022graphd} are being studied and utilized to tailor hardware designs specifically for graph-based reasoning tasks, offering new avenues for application-specific optimizations.

\item \emph{Real-time Learning Applications:}
The VSA computing paradigm is being extensively explored for its potential to deliver highly efficient learning capabilities, especially in real-time application scenarios. This effort has driven the development of hardware-friendly encoder~\cite{chen2022darl}, and non-MAC-based feature extractor with VSA classifiers, aimed at supporting low-latency retraining at the network edge~\cite{kandaswamy2022real}. To further reduce latency, some researchers have proposed completely bypassing the encoding computation by leveraging memory lookup techniques, offering a promising approach toward efficient real-time learning systems~\cite{imani2021revisiting}. 

\item \emph{General Applications:}
General hardware architectures have been proposed for a wide range of VSA applications, emphasizing reconfigurability and task-agnostic designs. These architectures span GPU/CPU~\cite{montagna2018pulp, menon2023accelerating}, FPGA~\cite{salamat2019f5}, ASIC~\cite{khaleghi2021tiny}, and IMC~\cite{garcia2022hdc8192, kazemi2022achieving}, providing flexible hardware design frameworks for implementing VSA algorithms.

\end{enumerate}
\begin{table}[t!]
\caption{Application-Specific Hardware Features Summarized from VSA Implementations.}
\small
\centering
\resizebox{\linewidth}{!}{%
\renewcommand{\arraystretch}{1.3}
\begin{tabular}{c|l|l}
\hline
    \multicolumn{1}{c|}{\textbf{Application}} & 
    \multicolumn{1}{c|}{\textbf{Reference Index}} & 
    \multicolumn{1}{c}{\textbf{Hardware Features}} \\
\hline
    \multirow{11}{*}{Classification}  & 
    [1] \cite{datta2019programmable} & 
    Flexible digital architecture, ROM + flip flops, DPU \\ 

    &
    [2] ~\cite{wu2018hyperdimensional} &
    3D integration of CNFET and RRAM, RRAM gradual reset \\
    & 
    [3] ~\cite{dutta2022hdnn} & 
    RRAM crossbar, Tile-based architecture  \\
    &
    [4] ~\cite{karunaratne2020memory} & 
    PCM-based, Partition method for AM search\\
    & 
    [5] ~\cite{morris2022stochastic} &
    Analog IMC, ADC-less \\
    &
    [6] ~\cite{wang2023computing} &
    Digital IMC, SRAM tile-based architecture, \\
    &
    [7] ~\cite{gupta2018felix} &
    Single-cycle IMC logic \\
    &
    [8] ~\cite{imani2019binary} &
    XOR array \\
    &
    [9] ~\cite{morris2021hydrea} &
    Adaptive bit-width change\\
    &
    [10] ~\cite{imani2019fach} &
    Two-stage pipeline, MAC hardware for clustered class vector \\
    &
    [11] ~\cite{fayza2023towards} &
    Electro-photonic accelerator for VSA training and inference \\

\hline
    \multirow{2}{*}{Genome Sequencing}  & 
    [12] ~\cite{hsu2022memory} & 
    High density 3D NAND flash memory \\ 
    &
    [13] ~\cite{kim2020geniehd} &
    Multi-platform compatibility, Reference buffer \\
\hline
    \multirow{5}{*}{Graph and Reasoning}  & 
    [14] ~\cite{poduval2022graphd} & 
    NOR-based, Switching characteristic of NVM  \\ 
    &
    [15] ~\cite{chen2023hypergraf} &
    Pipelined matrix multiplication, Domain-specific encoder and decoder \\
    &
    [16] ~\cite{barkam2023reliable} &
    FeFET-based, Learning realistic device noise \\
    &
    [17] ~\cite{kang2022relhd} &
    FeFET-based, Address sparseness and irregularity \\
    &
    [18] ~\cite{langenegger2023memory} &
    PCM stochasticity, In-memory MVM \\
\hline

    \multirow{3}{*}{Real-time Learning}  & 
    [19] ~\cite{kandaswamy2022real} & 
    Shift-ACCumulate instead of MAC, Neural network accelerator \\ 
    &
    [20] ~\cite{imani2021revisiting} &
    Map encoder module to simple memory lookup \\
    &
    [21] ~\cite{chen2022darl} &
    Hardware-friendly encoder IP, HV chunk fragmentation \\
\hline
    \multirow{5}{*}{Generic Architecture}  & 
    [22] ~\cite{montagna2018pulp} & 
    Processors cluster architecture \\ 
    &
    [23] ~\cite{garcia2022hdc8192} &
    Mixed-signal processing unit \\
    &
    [24] ~\cite{kazemi2022achieving} &
    FeFET-based content-addressable memory  \\
    &
    [25] ~\cite{khaleghi2021tiny} &
    Low-power, energy-efficient, low-latency ASIC platform \\
    &
    [26] ~\cite{menon2023accelerating} &
    Vector lanes in vector accelerator \\
    &
    [27] ~\cite{salamat2019f5}&
    DSP mapping for PE, Prefetch buffer for BRAM delay hiding\\
\hline
\end{tabular}
}
\label{table:hardware_map}
\end{table}

Summarized in Figure~\ref{fig:hardware-mappings}, generic architectures are positioned in the lower left corner due to their versatility across various application domains. The trend indicates that CPU/GPU-only, FPGA, and some IMC solutions offer moderate performance, striking a balance between latency and energy consumption. In contrast, a hybrid approach combining configurable ASICs/digital units with IMC can deliver superior performance despite its complexity. Such designs can be tailored for energy efficiency and latency reduction, creating a spectrum for exploring efficiency versus reconfigurability based on application needs. Additionally, the adoption of 2.5D and 3D chiplet integration technologies facilitates the seamless integration of diverse applications or modules, optimizing dataflow and enhancing computational efficiency.

\begin{figure}[!t]
	\centering
	\includegraphics[width=1.00\columnwidth]{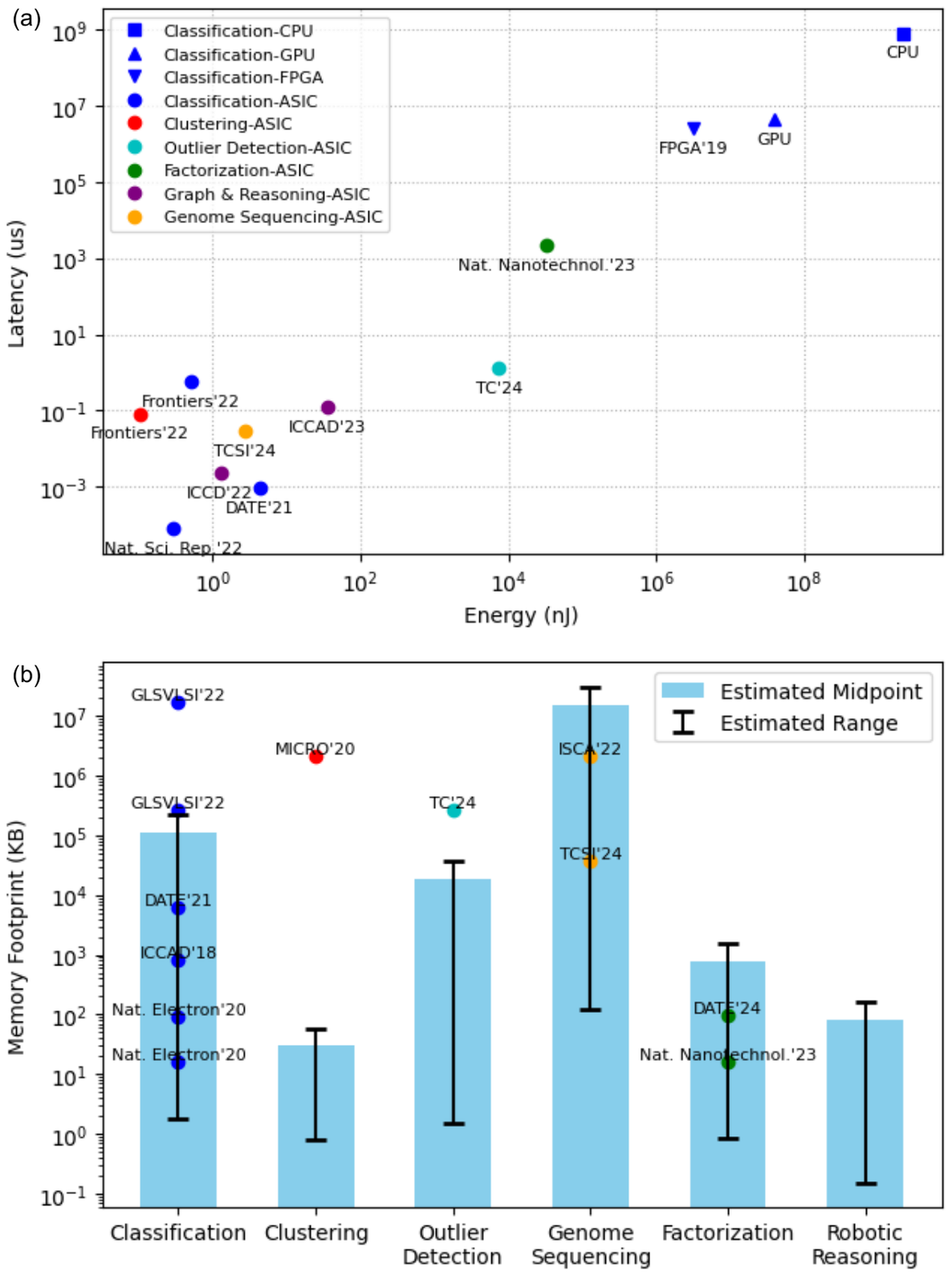}
        \caption{(a) The latency and energy comparison of VSA platforms in application domains. (b) Memory footprint range estimation (bars) of applications and memory usage from prior implementations (points).}
	\label{fig:application__latency-energy_memory}
\end{figure}

Based on the characteristics of hyperdimensional operations and the mechanism of different kernel implementations, we estimate the memory footprint bound of VSA applications (Figure~\ref{fig:application__latency-energy_memory}(b). In ASIC hardware architectures, the memory hierarchy includes both off-chip and on-chip memories that range from IMC units to buffers and register files. In our estimation, the configuration and usage of buffers and register files remain flexible, depending on the underlying dataflow strategy, while off-chip memory accesses can often lead to reduced system efficiency. Our methodology involves mapping operations that are feasible to be implemented in IMC cores. The IMC cores can also be used as pure storage for operands, aligning with prior IMC-based hyperdimensional computing designs that leverage IMC primarily for its high-density storage benefits. The factors that contribute to the difference in memory bound among applications include hypervector dimensionality, encoding methods, datasets, and function kernels on the algorithm side, as well as primitive designs and dataflow pipelining on the hardware side (further discussed in Section~\ref{sec:cognition}) . These factors differ across applications and require tailored design space exploration. Accordingly, we build on insights prior hardware implementations to guide our estimation of memory-bound performance.

The lower bound is estimated with a minimum memory requirement. At the algorithm level, the binary hyperdimensional representation and the encoding method that requires less memory capacity (random projection encoding) are used on the smallest dataset feasible for each application. At the hardware level, we adopt the state-of-the-art multi-bit PCM (3 bits per cell) to save the memory footprints. From the perspective of algorithm-hardware co-design, we evaluate the area-efficient accelerator designs while accounting for complexity and performance overhead. To estimate the upper bound, we consider the maximum memory requirement: using multi-bit hyperdimensional representation (4-bit in our case), the encoding method with the highest memory capacity requirement (n-gram encoding) on the largest dataset, and single-bit memory cells. We then analyze performance-efficient accelerator designs with the memory and area overhead factored in. The details of function kernels and their implementation and analysis for each application category are discussed below. 

Applications that can be categorized as classification, clustering, outlier detection, and genomics share the same fundamental kernels in the training and inference phase, i.e., encoding and similarity check. The logic implementation difference between these applications results in minor memory discrepancy. The factorization involves primitives of unbinding, similarity, and projection. The reactive robotic reasoning includes training and recall.

The minimum memory requirement for all applications is a few kilobytes, whereas the maximum order of magnitude varies from 100 KB to 1 GB. The clustering and robotic reasoning need less memory due to the simple encoding method in clustering and the smaller number of sensors and actuators used in robot reasoning. In contrast, classification and outlier detection demand more memory to store large-scale class datasets, with genomics requiring the most due to long sequence samples.
 
It is observed that the memory footprints of some previous works fall outside the estimated range. These deviations are attributed to specific design strategies that necessitated a departure from our generic estimation framework. For instance, \cite{dutta2022hdnn} factored in the feature extraction by applying matrix-vector multiplication in the dense neural network as the preprocessing step of the feature encoding in classification, \cite{imani2020dual} design with cluster parallelism to have higher performance, thus resulting in much higher memory resource usage. The multi-step in-memory permutation was developed in~\cite{wang2023computing}. These designs, while divergent, contribute to a broader understanding of the design space and demonstrate the flexibility of design strategies under varying constraints, prompting us to consider and analyze the arising trade-offs.

Suppose that the memory capacity is below the lower bound. In that case, the IMC cores may need reprogramming to handle changing operands, including inter- or even intra-kernel data traffic, which diminishes the benefits of parallel computing and reduced data movement between memory and compute units offered by IMC, ultimately leading to increased energy consumption and latency. Conversely, memory capacity exceeding the upper bound results in unnecessary area overhead and inefficiencies in energy and latency, particularly due to memory standby power or refresh cycles in the cases of charge-based memories.

\subsection{Unified Co-Design Framework}

\begin{figure}[!t]
	\centering
	\includegraphics[width=\columnwidth]{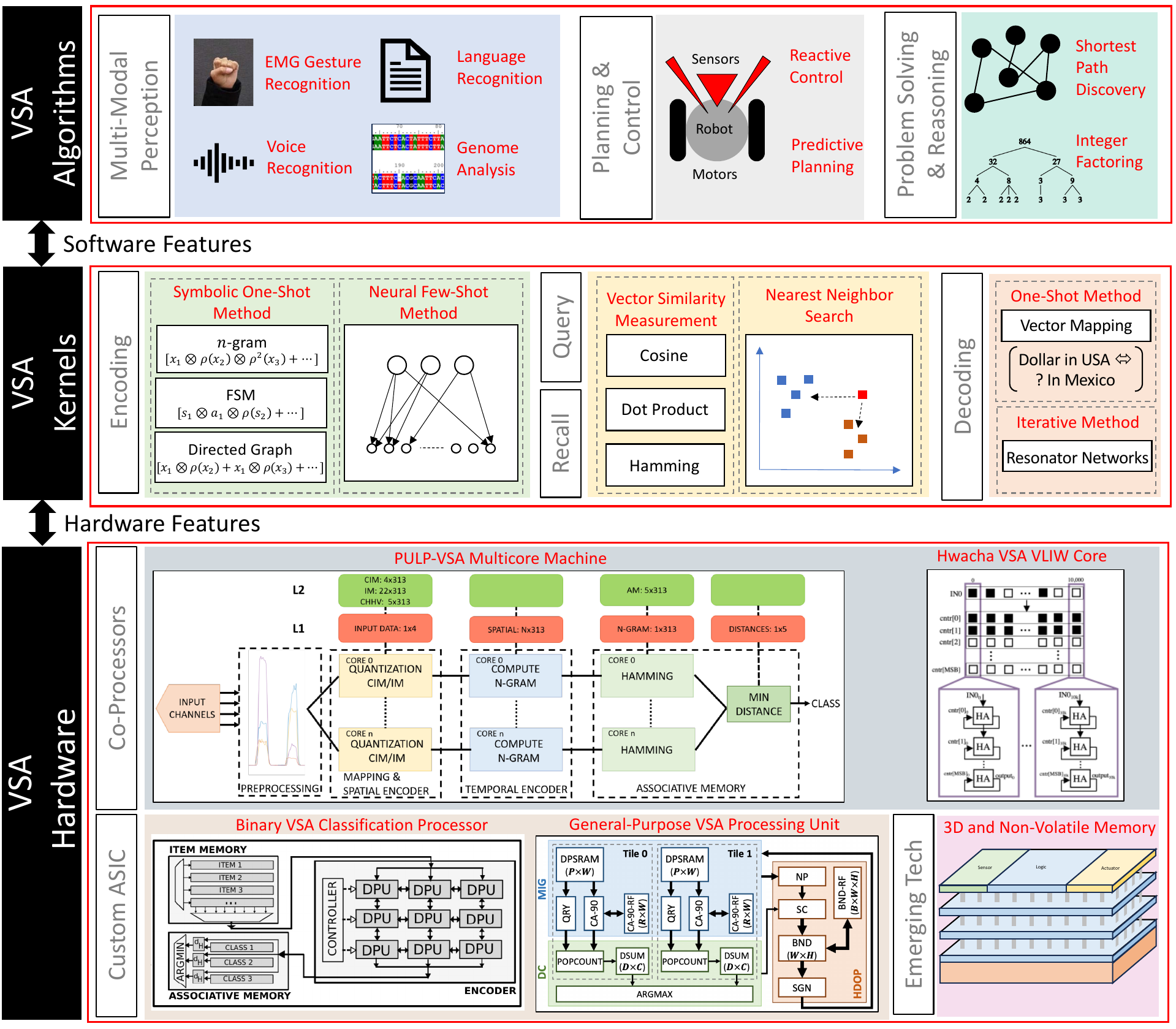}
	\caption{An illustration of the unified hardware/software co-design framework for VSAs.}
	\label{fig:VSA_codesign}
\end{figure}

The two co-design steps described above, \textit{kernel formulation} and \textit{feature-aware hardware realization}, can be approached independently. However, it is natural to think of these steps as being part of a unified co-design framework, bringing them into a merged design space for VSA-based cognitive systems. Figure~\ref{fig:VSA_codesign} illustrates the co-design framework that integrates both software and hardware perspectives for VSAs. It spans from high-level cognitive tasks, through kernel formulation, down to hardware architectures. The primary objective of this framework is twofold: (1) it allows holistic optimization across the entire design spectrum. A unified co-design framework endeavors to optimize not only the individual steps of kernel formulation and hardware realization but also their collective impact on the VSA system with respect to performance, latency, and energy. (2) It facilitates data-driven refinement of the design space, offering capabilities for dynamic adaptation to evolving requirements and technological landscapes. Through this co-design framework, hardware design methodologies for VSAs can be optimized, steering them towards a more systematic and top-down approach. In the following section, we explore an application case study that conceptualizes such a co-design framework.



\begin{figure}[!t]
	\centering
	\includegraphics[ width=1.00\columnwidth]{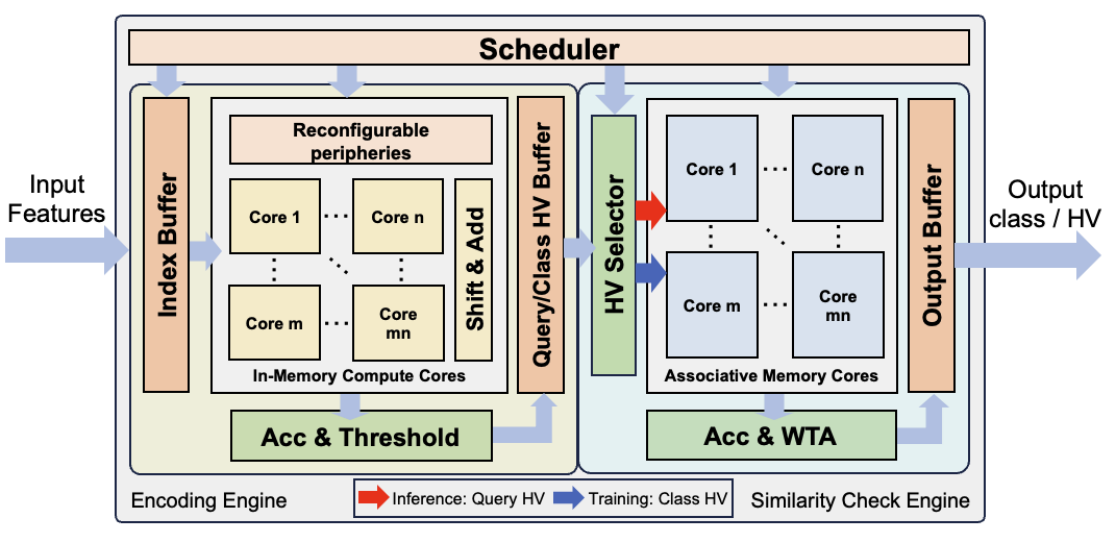}
        \caption{Generic in-memory compute paradigm for vector-symbolic architectures.}
	\label{fig:cim_template}
\end{figure}

Given the advantages of in-memory computing architectures in terms of reducing data movement and enhancing MVM computation efficiency, elaborated in Section~\ref{sec:hardware}, we develop a generic IMC reconfigurable architecture template (Figure~\ref{fig:cim_template}) as a synthesis and extension of state-of-the-art IMC-based VSA designs. This template is intended to be applied to complex VSA systems, enabling a flexible and reconfigurable approach that accommodates various precision, sparsity, and quantization requirements. The purpose of the template is to provide a hardware fundamental for the versatile framework that can be leveraged to develop and implement VSA systems more effectively. Using classification as the baseline application, the subsequent discussion elaborates on the architectural modules and their functionalities within this proposed template. The modifications and adaptations of the architecture to suit other applications are extended in Section~\ref{sec:cognition}.

\begin{figure}[!t]
	\centering
	\includegraphics[ width=.85\columnwidth]{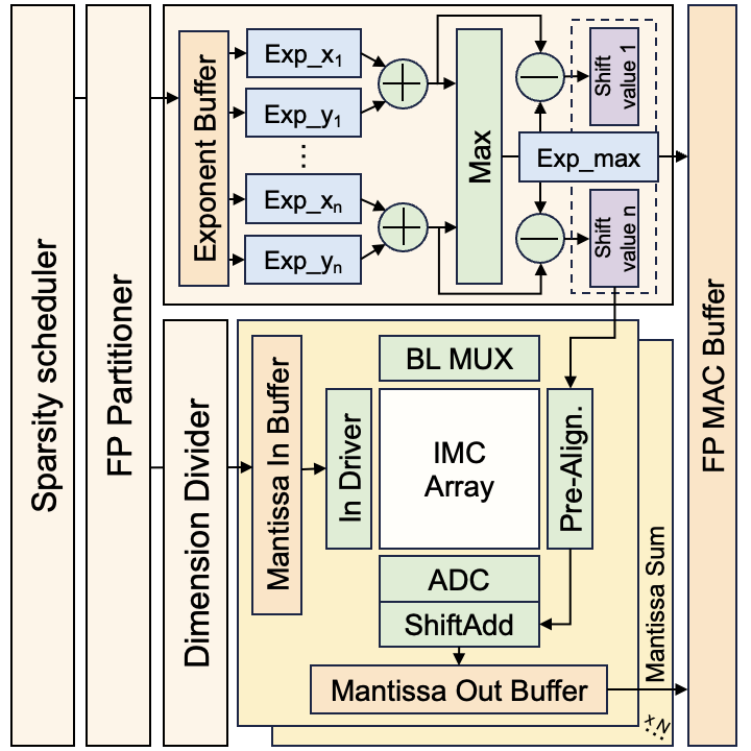}
        \caption{In-memory compute cores and peripheries for vector-symbolic architectures.}
	\label{fig:cim_peripheral}
\end{figure}

Our IMC architecture comprises two primary engines: the encoding engine and the similarity check engine. The encoding engine includes IMC cores, reconfigurable peripheries, adder and threshold logic, controller, and buffers, all of which enable high-dimensional computation. The IMC cores implement in-memory binding and bundling. Reconfigurable peripheries handle varying data precision (i.e., floating point or integer values), sparsity levels, and vector lengths, which are crucial for supporting diverse VSA workloads in IMC architectures. In particular, the sparsity scheduler enables efficient processing by skipping zero-valued computations and dynamically allocating resources based on sparsity patterns, an approach that aligns with prior work on sparse-aware CIM architectures~\cite{dai2019graphsar,giannoula2022sparsep,xie2021spacea}. Adder logic circuit aggregates the hypervectors for each class and adds (subtracts) the hypervector to (from) class hypervectors in the training phase. Threshold logic supports various scenarios, such as generating the binary class hypervectors from the summed multiple binary feature hypervectors or generating similarity metrics. Buffers in the encoding engine are partitioned to store item memory for input features, and to store generated query and class hypervectors. IMC peripheries (Figure~\ref{fig:cim_peripheral}) include sparsity scheduler, FP (floating-point) partitioner that separates mantissa and exponent, and a dimension divider for handling high-dimensional computations. The exponent computing module facilitates exponent and mantissa processing and integrates seamlessly with the IMC cores, enabling efficient floating-point computation acceleration. The floating-point MAC implementation includes exponent alignment and addition, and mantissa multiplication. First, the exponents are read out from the exponent buffer and added in pairs based on the mechanism of MAC implementation, and the maximum sum value among all pairs is determined. The exponent shift value for each pair is calculated by subtracting the maximum sum from all exponent sums. These shift values align the exponent part and adjust the mantissa product generated by the IMC cores. The final floating-point MAC result is obtained by integrating the aligned mantissa sum with the previously determined maximum exponent.

The similarity check engine includes the following components – associative memory cores, controller, adder and winner-takes-all logic (or comparator logic), and buffers – which enable the class labeling after computing the similarities. The class hypervectors are updated in associative memory cores in the inference phase, prestored and fixed for the inference phase. Associative memory cores can be either in-memory computing cores or associative-memory cores depending on the similarity computation method. With the adder and comparator logic, the engine calculates the similarity between the query hypervector and each class hypervector by accumulating matching bit values in all dimensions. The class label of the query hypervector is determined and output based on the comparison result.

\section{Case Study for Hierarchical Cognition}
\label{sec:cognition}

Insights from the field of cognitive science have suggested that humans build complex models of the world to fulfill the fundamental tasks of cognition~\cite{jeon2014hierarchical}. Cognitive workloads, which span from low-level perceptual processes that integrate massive sensory data from diverse sources to high-level problem-solving and reasoning, are inherently both heterogeneous and hierarchical. These two characteristics motivate the selection of hierarchical cognition as a representative case study in this work. First, hierarchical organization mirrors the intricate layers of cognitive processing, forming a continuum that orchestrates our interactions with the environment. From a computational perspective, the research literature includes many VSA methods that can capture the underlying mechanisms of these cognitive processes. These methods collectively form a unified cognitive framework, which stands to benefit several application domains, in particular assistive technologies and autonomous systems~\cite{dumont2023exploiting, menon2022brain, olascoaga2022brain}. Figure~\ref{fig:VSA_codesign} illustrates the unified co-design framework with the case study of hierarchical cognition. Second, given the diverse nature of the applications and kernel operations involved in the cognition system, the corresponding hardware designs must also be tailored accordingly. Building on this, we present the first in-memory computing hierarchical cognition hardware system as a design example based on our IMC template introduced in Section~\ref {sec:codesign}. 
The following subsections provide a brief overview of the relevant VSA methods, including their fundamental kernels, their implementations within the IMC template, and key hardware design considerations from functional partitioning, dataflow mapping, to memory type selection, and the impact of technology node scaling.

\subsection{Multi-Modal Perception}
\label{subsec:perception}


An increasing number of sensor types (such as gyroscopes, accelerometers, and EEGs) are integrated into one edge device to detect and collect richer data in the environment. This multimodal nature of data presents challenges in processing and learning, specifically, how to exploit multi-modality to facilitate better learning and how to efficiently process large-scale data on resource-constrained embedded devices. VSA-based architecture unlocks potential solutions to efficiently learn from multimodal data at the edge~\cite{zhao2023attentive,chang2019hyperdimensional,menon2021highly}. The multimodal data from different sensor types are first encoded into hypervectors with uniform dimensions by utilizing the permutation-based encoder, and then the hypervectors of different modalities can be either bundled or processed through attention modules to learn intermodality correlations for downstream perceptual tasks.

VSA-based perception methods can also build on data structures such as graphs, inspired by the human brain that clusters data and represents information as a graph structure~\cite{bassett2017network}. The objects and edges in the graph can show the correlation between objects, and memorization provides prior knowledge to keep the context and define confidence for downstream reasoning and decision-making. Specifically, high-dimensional vectors can be utilized to holographically represent the nodes and memorize the graph, which enables the known information (such as graph nodes and their connections) to be well memorized~\cite{poduval2022graphd,ma2018holistic,nickel2016holographic}. 
For a graph with $V$ nodes and $E$ edges, the node $i$ memory is constructed by accumulating all node hypervectors connected to it as $M_i = \sum_j H_j$ where $H$ is a random hypervector to the node and $j$ represents all the neighbors of node $i$. The bundling of all associated hypervectors then generates a graph memory as $G = \sum_i H_i \oplus M_i$, where the graph memory is a compressed, invertible, and transparent model that can be used for downstream brain-like cognitive learning tasks.

VSA-based architecture also excels in cognitive perception when confronted with out-of-distribution (OOD) samples - an inevitably encountered scenario in many embodied AI applications, such as robotics and autonomous systems, when objects and scenes were not part of the training data distribution. The powerful concepts of binding, bundling, and projection from VSAs can be applied to the features from multiple layers in a DNN without requiring re-training or any prior knowledge of the OOD data. Hyperdimensional Feature Fusion (HDFF) presents such an example~\cite{wilson2023hyperdimensional}.
Specifically, the feature maps from multiple layers can be projected into a common vector space by using similarity-preserving semi-orthogonal projection matrices. During deployment, the projection and bundling operations are applied for a new input image, i.e., $y = h_1 \otimes h_2 \cdots \otimes h_l$ where $h_i$ is the encoded high-dimensional vector from different network layers and  $y$ is the resulting single vector that serves as an expressive descriptor for the input image. The cosine similarity is then leveraged by the class representatives to identify OOD samples. The VSA-inspired technique paves the way to potentially address the limitations of DNN models in OOD scenarios, where they tend to fail silently in producing overconfident but erroneous predictions.

\begin{figure}[!t]
	\centering
	\includegraphics[width=\columnwidth]{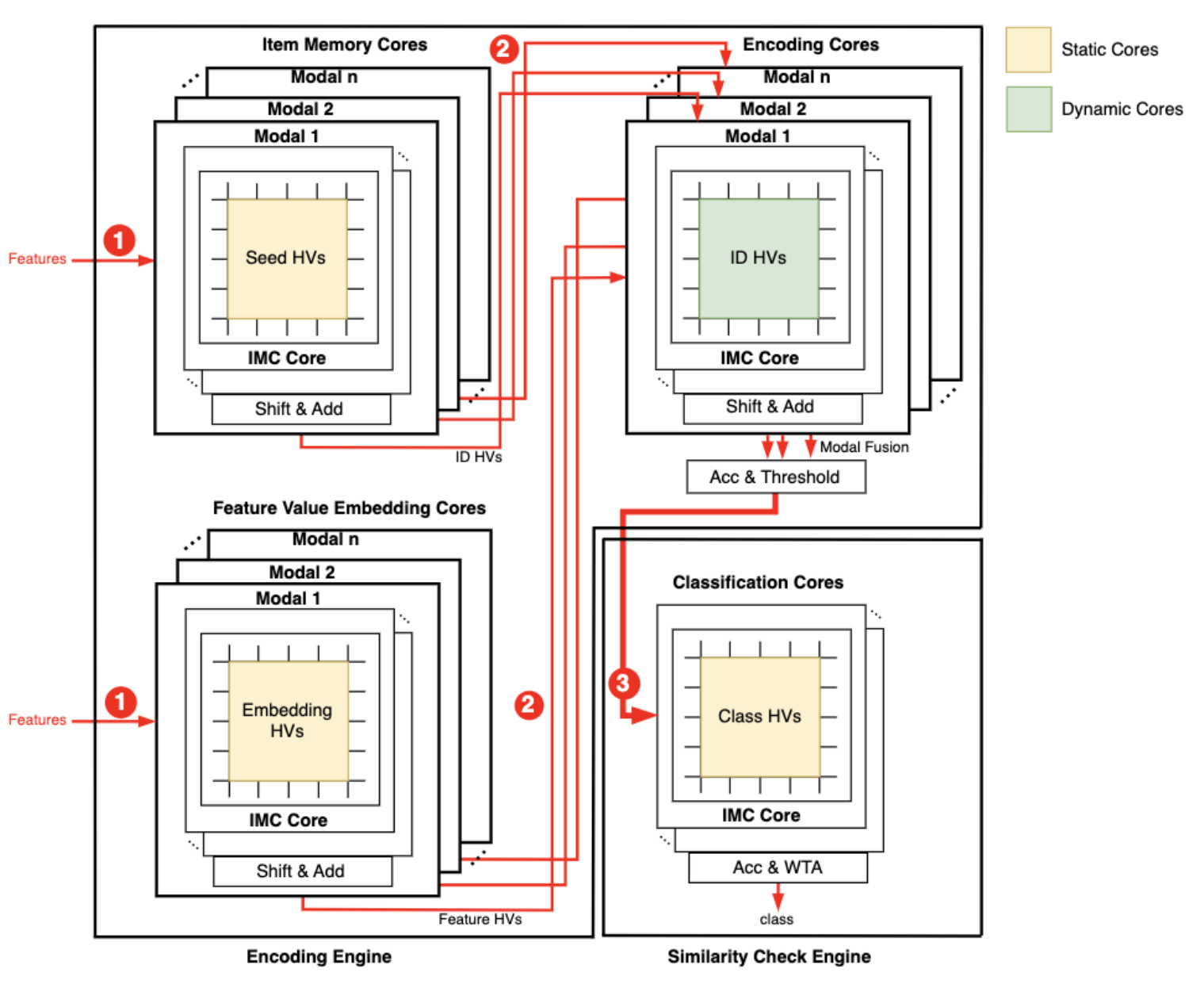}
	\caption{An illustration of the IMC multi-modal perception kernel design.}
	\label{fig:multimodal_cim}
\end{figure}

IMC-based hardware implementation leverages diverse memory characteristics to optimize different cores. Figure~\ref{fig:multimodal_cim} illustrates an IMC multi-modal perception kernel design. The Feature Value Embedding Cores, Item Memory Cores, and Classification Cores are static cores; all memory cells are prestored, eliminating the need for additional memory writes during fusion operations. In contrast, Encoding Cores operate as dynamic cores as they process changing feature value/ID hypervectors as inputs. To illustrate the computation process, first, the features captured by modal sensors are mapped into feature value hypervectors in Feature Value Embedding Cores $\normalize{\text{\textcircled{\scriptsize 1}}}\normalize$, where in-memory random projection is used in this case as an example. The IDs of the features are also mapped to ID hypervectors in Item Memory Cores $\normalize{\text{\textcircled{\scriptsize 1}}}\normalize$, which serve as the in-memory look-up table. Each modal's generated feature value hypervectors and feature ID hypervectors are encoded into modal hypervectors through in-memory binding and bundling in Encoding Cores $\normalize{\text{\textcircled{\scriptsize 2}}}\normalize$. A temporal encoding step might be needed to create a hypervector of each modality according to the features capturing process, which can be realized by the permutation function of shift-and-add logic in the template. As the final step in the Encoding Engine, the modal fusion is implemented through near-memory bundling of each modal hypervector in Encoding Cores. In Similarity Check Engine, the fused hypervector is bundled/compared with class hypervectors in Classification Cores $\normalize{\text{\textcircled{\scriptsize 3}}}\normalize$ in the training/inference phase.

\subsection{Planning and Control}

While pattern recognition is the most popular application of VSAs, there is a wide body of research showing that VSAs can also facilitate planning and control applications. Within this domain, VSA methods can be broadly classified into two primary groups: \emph{reactive} and \emph{predictive}. A reactive VSA method allows an agent to run task demonstrations while encoding its interactions as pairs of sensor input and actuator output~\cite{neubert2019introduction, mitrokhin2019learning}. The so-called reactive robotic behavior is eventually encoded as a single high-dimensional vector, thus enabling an agent to promptly respond (through a quick memory search) to immediate stimuli during navigation~\cite{Menon2022, menon2023shared}. A predictive method, on the other hand, aims to learn an abstract view or a topology of the environment's state space. Remarkably, VSAs have demonstrated great strengths in representing multiple data structures~\cite{kleyko2022vector}, and hence they can be used to learn structural, most often graph, representations of an environment~\cite{ni2022hdpg, stockl2024local}. Such a VSA representation is often referred to as \emph{a cognitive map}~\cite{mcdonald2023modularizing}. Additionally, planning and control are usually vulnerable to software or hardware disturbance~\cite{hsiao2023silent,wan2022analyzing}, thus, the robustness characteristics of VSA computing benefit the safety of autonomous agents.

\begin{figure}[!t]
	\centering
	\includegraphics[width=\columnwidth]{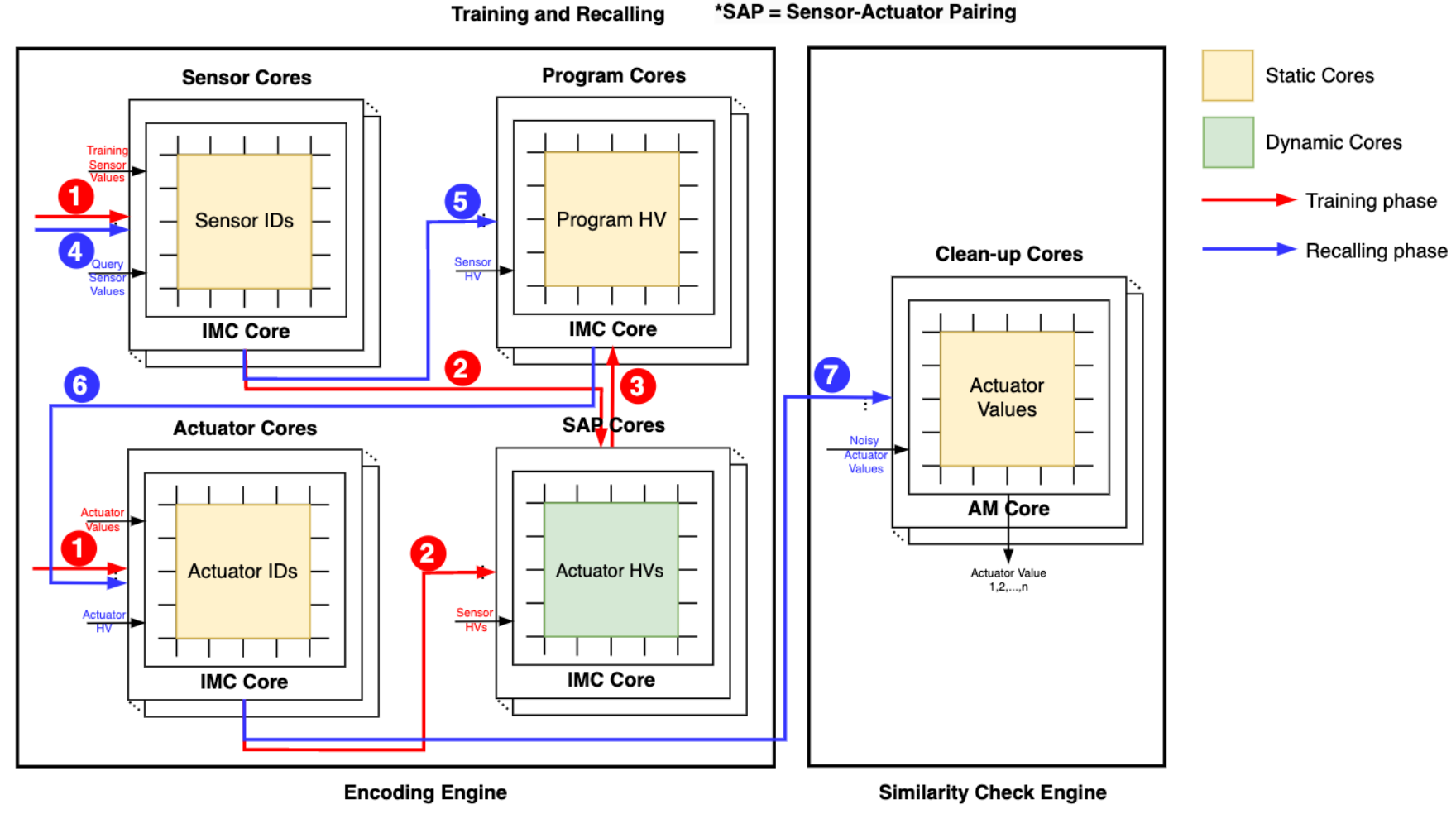}
	\caption{An illustration of the IMC kernel design in an example of reactive robot navigation tasks.}
	\label{fig:plan_control_cim_v3}
\end{figure}

An IMC kernel design is demonstrated through an example of reactive robot navigation tasks~\cite{neubert2019introduction} (Figure~\ref{fig:plan_control_cim_v3}), utilizing static and dynamic cores to optimize different operations. In the 2-D navigation task, each of the sensors and the actuators (for the robot movement direction in this case) is assigned a random ID hypervector, which is bound with input/output feature value hypervectors to form sensor-action pairs, enabling the system to learn for future tasks. The task includes a training phase and a recalling phase. During training, the in-memory hyperdimensional operations are only implemented in the encoding engine, including the Sensor Cores, Actuator Cores, and Program Cores that handle operations with static memory cores mapping, and the dynamic memory cores, i.e., Sensor-Actuator Pairing (SAP) Cores where both inputs and memories change with training samples. The training phase starts with encoding each sensor value into a sensor hypervector through in-memory binding (XOR) followed by in-memory bundling (addition) operations to generate a single sensor hypervector representing the training sample, where ID hypervectors for all sensors are pre-stored in the Sensor Cores $\normalize{\text{\textcircled{\scriptsize 1}}}\normalize$. A similar process is applied to actuator data in the Actuator Cores. Thus, the sensor-actuator pairs of each training sample are created. For each pair, the sensor hypervector (input) is bound with the corresponding actuator hypervector (stored in memory). The sensor-actuator pairs from all training samples are bundled into the program hypervector in SAP Cores $\normalize{\text{\textcircled{\scriptsize 2}}}\normalize$. The generated program hypervectors are stored in Program Cores memory, preparing for the recalling phase $\normalize{\text{\textcircled{\scriptsize 3}}}\normalize$. 

During the recalling phase, both the encoding engine and the similarity check engine are mapped with pre-stored static memories, where some memories are reused from the training phase. The same encoding process is performed on the query input sensor data in Sensor Cores $\normalize{\text{\textcircled{\scriptsize 4}}}\normalize$. The query sensor hypervector is bound to the trained program hypervector pre-stored in Program Cores $\normalize{\text{\textcircled{\scriptsize 5}}}\normalize$. The generated actuator hypervector goes through the unbinding with actuator ID hypervectors in Actuator Cores, resulting in a noisy actuator hypervector for each actuator $\normalize{\text{\textcircled{\scriptsize 6}}}\normalize$. These actuator hypervectors undergo the clean-up process in the Similarity Check Engine by comparing against the stored hypervector representations of possible robot actuation values in Clean-up Cores $\normalize{\text{\textcircled{\scriptsize 7}}}\normalize$. The actuation with the highest similarity to the noisy actuator hypervector is provided as the action output (move in one of the directions in this case) to the robot.

\subsection{Problem Solving and Reasoning}
\label{subsec:res_net}

VSAs are key enablers for neuro-symbolic methods, offering highly efficient solutions for a range of reasoning and analogy-making problems~\cite{emruli2013analogical, hersche2023probabilistic}. Notably, VSA methods for reasoning can be broadly classified into two groups: \emph{rule-based} and \emph{disentanglement-based}. The objective of rule-based methods is to extract relevant information from a composite vector and transfer it to unfamiliar contexts where the original items may not be known apriori. Consider an example that involves the following rules: 
\[ 
States=[name\otimes USA + cur\otimes DOL],
\]
\[ 
Mexico=[name\otimes MEX + cur\otimes PES],
\]
\[
F=States\otimes Mexico
\]
A typical scenario for rule-based reasoning is when we need to find an answer to the question: ``\emph{What is the Dollar of Mexico?}''~\cite{kanerva2010we}, which is formulated as a function $DOL\otimes F$. This function computes the reverse mapping of $F$, aiming to find what in $Mexico$ corresponds to $DOL$. The returned vector, however, is a distorted version of the correct answer ($PES$) because of the cross-talk noise coming from all other elements. The correct answer can be recovered in turn using an implementation of the clean-up memory model. The described approach can be generalized and hence applied to more complex scenarios, such as Raven's progressive matrices~\cite{hersche2023neuro} and analogical mapping through graph isomorphism~\cite{gayler2009distributed}. 


One more challenging scenario arises when reasoning involves multiple unknown factors. For instance, suppose we have a composite vector $f=a\otimes b\otimes c$, which is defined by three unknown factors: $a$, $b$, and $c$. These factors are unknown because we only have access to the codebooks ($A$, $B$, and $C$), which represent the high-dimensional item sub-spaces for these factors. To recover the original items of $f$ using the rule-based approach described earlier, we would need to search through all possible combinations of the factors, which becomes increasingly complex as the number of factors grows. In such scenarios, the disentanglement-based approach (resonator networks) comes into play~\cite{frady2020resonator}. 

The disentanglement-based reasoning approach formulates a compositional state-space model, which superposes all possible combinatorial solutions. Given this model, distributed disentanglement computations are performed while searching for the exact feature or item vectors. One could employ this approach to disentangle $f$, defined earlier, into its factors. The state-space equations that describe this method are given as follows: 

\[
\hat{a}(t+1) = g\big(AA^\top(f\otimes \hat{b}(t)\otimes \hat{c}(t))\big);\ \ A=[a_1\ a_2\ \ldots]
\]\[
\hat{b}(t+1) = g\big(BB^\top(f\otimes \hat{a}(t)\otimes \hat{c}(t))\big);\ \ B=[b_1\ b_2\ \ldots]
\]\[
\hat{c}(t+1) = g\big(CC^\top(f\otimes \hat{a}(t)\otimes \hat{b}(t))\big);\ \ C=[c_1\ c_2\ \ldots]
\]

\noindent Here, $g(\cdot)$ is the sign function; $t$ is a time step; $\hat{a}$, $\hat{b}$, and $\hat{c}$ hold the predicted values of the factors $a$, $b$, and $c$, respectively. The term $AA^\top \hat{x}$ represents the update role for $\hat{a}$, which computes the dot product between the reverse-mapping vector $\hat{x}$ and the item vectors of $A$, scales these item vectors using the dot-product (weight) results, then bundles all the weighted vectors to update the state.

\begin{figure}[!t]
	\centering
	\includegraphics[width=\columnwidth]{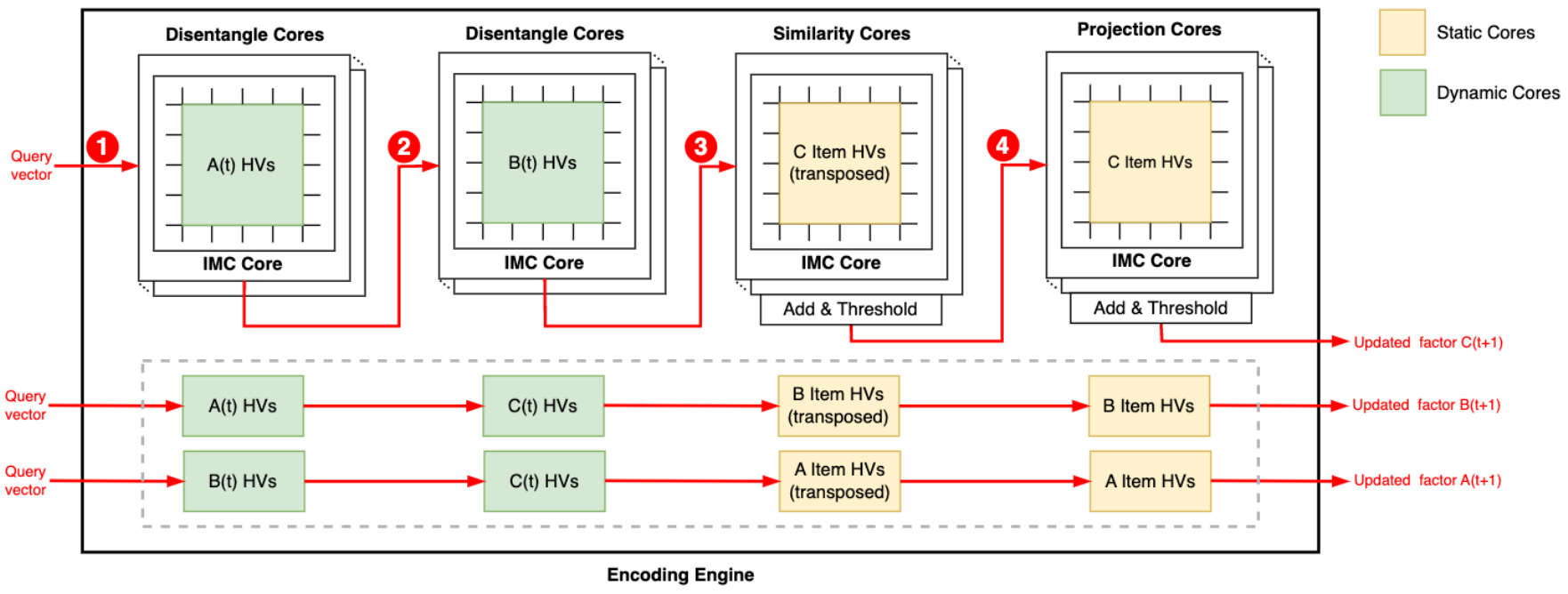}
	\caption{An illustration of the IMC reasoning kernel design in an example of a three-feature factorization task. The operations are partitioned into static and dynamic cores in the parallel factoring scenario.}
	\label{fig:problem_solve_cim_v2}
\end{figure}

Figure~\ref{fig:problem_solve_cim_v2} illustrates an IMC reasoning kernel design in an example of a three-feature factorization task. The query hypervector is made up of the characteristics from all features, resulting in the factoring for each feature. Take feature C factorization as an example, the query vector first unbinds the features A and B in sequential by undergoing in-memory binding in Disentangle Cores ($\normalize{\text{\textcircled{\scriptsize 1}}}\normalize$-$\normalize{\text{\textcircled{\scriptsize 2}}}\normalize$). The unbinding result goes through Similarity Cores and Projection Cores to implement the dot product with C feature codebooks and get factor C updated in this iteration ($\normalize{\text{\textcircled{\scriptsize 3}}}\normalize$-$\normalize{\text{\textcircled{\scriptsize 4}}}\normalize$). Multiple iterations are needed until each feature converges to the correct factorization with a high similarity for a code vector. In this task, factorization of each feature can either be handled sequentially or in parallel, where the IMC cores mapping and overall memory resources utilization vary substantially. 

In the case of sequential factoring, at each time step, the estimation update begins with the first factor. Subsequently, the second factor is updated using the newly estimated first factor and the most recent estimates of the remaining factors for the following similarity computation and projection process. This update process continues iteratively, with each subsequent factor updated based on the most current estimates available for all preceding factors. During the factorization, Disentangle Cores, similarity Cores, and Projection Cores are all dynamic cores. The feature types and their feature factors estimation stored in disentangle cores change with the factoring iteration, and the stored feature codebooks for similarity and projection steps vary depending on the feature type in each factoring iteration. In the case of parallel factoring (Figure~\ref{fig:problem_solve_cim_v2}), all features are updated at the same time step and are mapped in cores for in-memory binding for the next iteration. The memory footprint is scaled with the number of features. While the Disentangle Cores are dynamic cores, storing updated feature factors estimation with fixed feature types, the Similarity Cores and Projection Cores are static cores, with each feature codebook written once into their respective static cores. 

\subsection{Hardware Realization}
\label{subsec:app_hw_mapping}

Developing a hardware platform for the above framework involves careful consideration of the requirements of each layer. This can be achieved by exploring the kernels, computational characteristics, and constraints of each layer and examining the hardware technologies that are aligned with them. 

\begin{figure*}[!t]
	\centering
	\includegraphics[width=\textwidth]{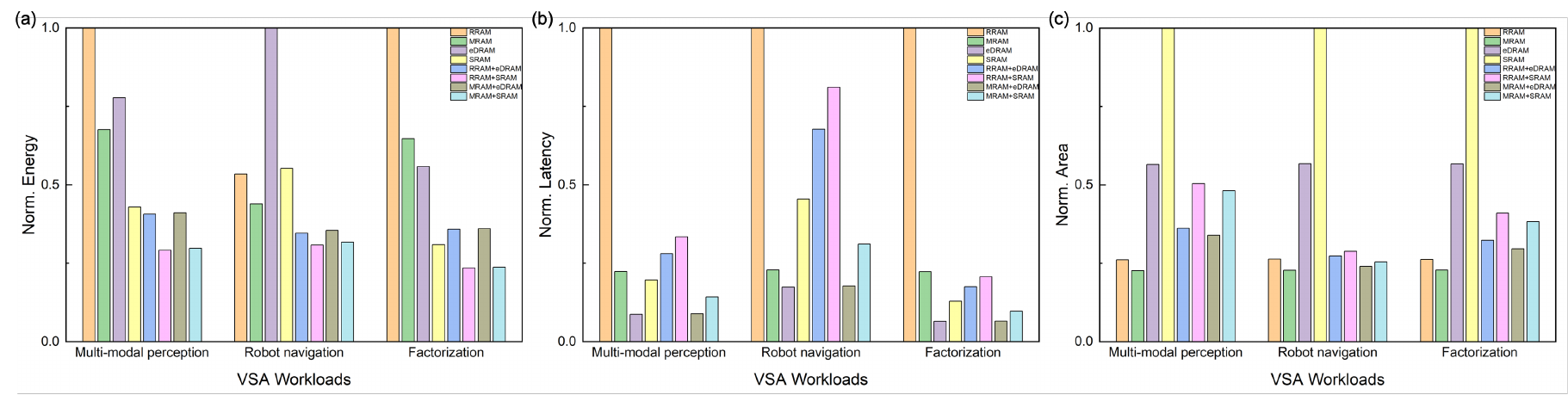}
	\caption{Normalized energy, latency, and area benchmark across IMC architectures of different 65 nm memories for VSA workloads. The heterogeneous charge-based memory and NVM designs improve the total energy and latency compared to the single-memory IMC baseline.}
	\label{fig:memory_benchmark}
\end{figure*}

\begin{figure}[!t]
	\centering
	\includegraphics[width=\columnwidth]{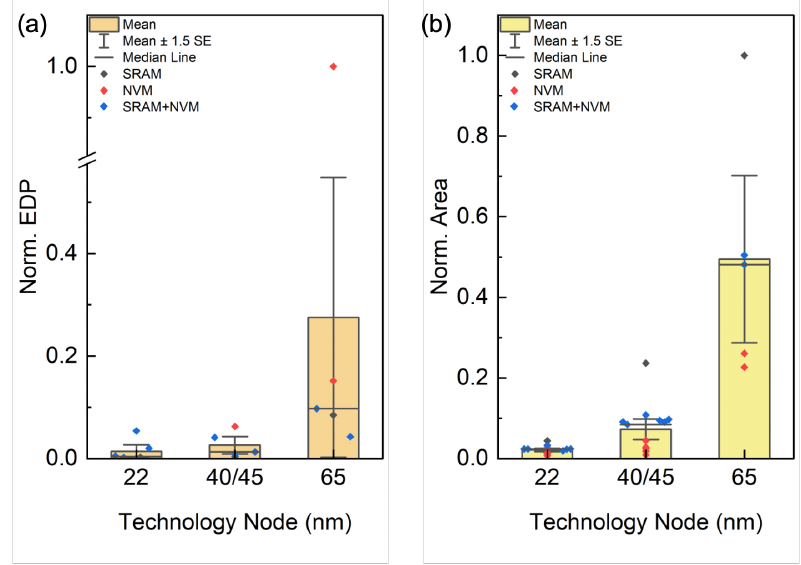}
	\caption{Normalized energy-delay product (EDP) and area benchmark across technology nodes (from 22 nm to 65 nm) of different memories for multi-modal perception workload.}
	\label{fig:memory_tech_benchmark}
\end{figure}

IMC-based architectures offer a natural solution due to their ability to minimize data movement and provide efficient in-memory vector operations. As introduced in Section~\ref{subsec:perception}, IMC kernel design tailored for multi-modal perception leverages static and dynamic memory cores to optimize different operations: Static cores precompute hypervectors, eliminating the need for additional memory writes during fusion operations; Dynamic cores handle feature value/ID hypervectors as inputs, supporting real-time encoding and temporal processing through in-memory operations. We evaluated the energy, latency, and area overhead of IMC architectures that integrate static and dynamic cores across three cognitive VSA applications (Figure~\ref{fig:memory_benchmark}) and the impact of technology scaling (Figure~\ref{fig:memory_tech_benchmark}). 

 Our analysis considers various memory core options, including charge-based memories~\cite{sun2009novel,choi2015refresh,wang202040,5280731,xu2015impact} (e.g., eDRAM and SRAM) and non-volatile memories ~\cite{chang2013low,spetalnick2024edge,giordano2021chimera,mu20241mb,wang202122,choe2023recent,oka20213d,lu2015fully,he2017exploring,agrawal2018spare,jain201913,hung2021four,xue2021cmos,chih202013,chiu202322nm,chang202013,shimoi202322,naik2020jedec,wei201913,zhang202222nm,rossi2021vega} (e.g., RRAM and MRAM). As shown in Figure~\ref{fig:memory_benchmark}, diverse memory types can be applied homogeneously across all IMC cores—either dynamic or static—or heterogeneously, with charge-based memories used in dynamic cores to efficiently handle frequent memory writes while mitigating standby energy overhead by minimizing SRAM leakage time and eDRAM refresh cycles, and NVMs in static cores to reduce write overhead and enhance energy-area efficiency. In architectures with heterogeneous memory cores, applying charge-based memories and NVM to dynamic and static IMC cores, respectively, results in relatively lower energy consumption compared to homogeneous memory configurations. Among the evaluated workloads, multi-modal perception and factorization have the worst energy efficiency when using RRAM, primarily due to frequent writes in dynamic cores and the inherently high write overhead of RRAM. Robot navigation performs worst with eDRAM, as the frequent refresh operations and retention time required during training across multiple samples incur significant energy overhead. Heterogeneous memory cores also demonstrate higher area efficiency compared to using charge-based memories alone, owing to the comparable dynamic and static IMC core memory footprint requirements and high-density nature of NVM.

As described in Section~\ref{subsec:perception}, multi-modal perception involves a computational process of continuous data sampling and HD encoding, with data extracted from a large number of sensory inputs. It is therefore logical to think of multi-modal perception as a time-critical application where data needs to be encoded in a timely manner~\cite{menon2021highly}. In other words, a hardware platform developed to realize multi-modal perception needs to be positioned within the top part of Figure~\ref{fig:hardware-mappings}. As shown in Figure~\ref{fig:memory_benchmark}, heterogeneous MRAM/eDRAM and MRAM/SRAM architectures achieve over 86\% lower latency compared to RRAM while maintaining latency performance close to eDRAM, which offers the lowest latency. Additionally, these configurations improve energy efficiency by up to 2.6× compared to eDRAM, balancing both speed and power effectively.

For the reasoning layer (Section~\ref{subsec:res_net}), NVM technologies play a crucial role in maintaining the persistent representations of knowledge and providing technology-assisted robustness to the resonator networks. These technologies are particularly well-suited for reasoning tasks due to their energy efficiency, non-volatile nature, and fast access times~\cite{langenegger2023memory}. Moreover, these memory technologies can be organized hierarchically (e.g., via 3-D stacking) to mirror the structure of the reasoning kernels, allowing for quick retrieval of stored information during iterative reasoning processes~\cite{wan2024h3dfact}. IMC architectures offer an intrinsic advantage by reducing data movement and enabling energy-efficient vector operations. Combining NVM and IMC enables parallelized vector operations, low-energy iterative updates, and scalable memory management, ensuring efficient and robust execution of reasoning tasks. Overall, reasoning computations can be realized with hardware platforms that are positioned at the bottom right part of Figure~\ref{fig:hardware-mappings}, indicating that reasoning kernels are not time-critical, and energy efficiency remains a key optimization goal. In this context, heterogeneous MRAM/SRAM and RRAM/SRAM architectures are well-suited for low-power requirements, reducing energy consumption by over 76\% compared to RRAM. At the same time, these architectures maintain performance comparable to eDRAM, the lowest-latency option.

In Figure~\ref{fig:memory_tech_benchmark}, energy-delay product (EDP) and area are benchmarked across 22 nm, 40/45 nm, and 65 nm technologies using different memories under the multi-modal perception workload. As the node goes from 65 nm to 22 nm, the average EDP and area overhead across the three memory configurations, SRAM alone, NVM alone, and hybrid SRAM/NVM, decrease significantly, aligning with the expectations of technology scaling. Hybrid SRAM/NVM and SRAM-only architectures exhibit improved EDP compared to NVM-only counterparts, primarily due to the mitigation of NVM write overhead. However, this advantage diminishes at more advanced technology nodes (22 nm and 40/45 nm), where the EDP performance of hybrid SRAM/NVM and SRAM-only designs becomes comparable. This trend is likely attributed to the fact that technology scaling primarily benefits silicon-based components, whereas NVM devices exhibit limited scalability with advanced technology nodes. These observations suggest that adopting advanced technology nodes in VSA kernel implementations can yield system-level energy and performance gains, but the limited scalability of NVM devices with technology nodes should be carefully considered during early-stage design exploration at the architecture level, with awareness of technology constraints. In addition, all configurations involving NVM—including NVM alone and hybrid SRAM/NVM—consistently demonstrate superior density across all technology nodes. This makes hybrid charge-based memories and NVM particularly promising candidates for energy-efficient and compact integration in edge VSA systems, where minimizing footprint and maximizing memory capacity are critical.

However, it is equally important that the developed hardware platform remains reconfigurable to support different encoding schemes for sensory data~\cite{datta2019programmable}. A hybrid approach that integrates IMC-based computation with digital FPGA or ASIC implementations would enhance both computational efficiency and adaptability across various applications. At the same time, advancements in memory technologies introduce additional considerations. For instance, state-of-the-art refresh-free eDRAM and ultra-low-power SRAM designs have the potential to further optimize power and performance trade-offs. As memory technologies continue to evolve, the optimal choice of memory for different applications may shift, reinforcing the need for flexible hardware platforms. In this context, digital processors based on FPGA or ASIC implementations would be particularly beneficial in maintaining reconfigurability while adapting to emerging memory innovations.

\section{Research Challenges and Opportunities}
\label{sec:challenge}

This section discusses major challenges for VSA system design and highlights opportunities for future research directions.

\subsection{Modular Formulation of VSA Kernels and Features}
The field of VSA computing is fast-changing, and thereby we foresee a significant increase in the number of newly proposed algorithms in the coming years~\cite{kleyko2023survey}. Being an algebra, VSA algorithms intrinsically share similar computational features and statistical properties, allowing designers to build libraries or packages of VSA functions~\cite{heddes2023torchhd}. It is therefore natural to seek a modular approach when formulating VSA kernels. Adopting this approach requires an optimization framework that assembles VSA kernels hierarchically in a plug-and-play fashion. This approach opens the door for \emph{hardware-aware} exploration of VSA methods without greatly delving into details of hardware design. Another key benefit of adopting a modular design approach is the capability to adapt and reconfigure deployed VSA kernels into unseen scenarios and algorithms. This dynamic adaptability can eventually be transformed into a real-time reconfiguration framework---akin to an ``operating system'' tailored specifically for VSA methods.

\subsection{Pareto-Front Exploration of Optimized Hardware Mapping and Compositional VSA Modeling}
Compositionality is one of the key characteristics of VSA methods. For instance, attribute vectors resulting from multiple tasks (e.g., perception and reasoning) can be bundled/superposed to generate more complex high-dimensional structures in a hierarchical fashion~\cite{kortylewski2019greedy}. This provides potential for a generic hardware design with maximally reusable modules. On the other hand, an efficient hardware design seeks to facilitate seamless data transfer devoid of contention across all abstraction levels, spanning from individual units to the system-on-chip level. Hence, while the computing-in-superposition nature inherent in VSAs creates massive opportunities for theoretical explorations, it may also pose challenges in terms of distributed processing and contention-free utilization of critical memory resources. Therefore, an efficient strategy for hardware mapping of VSA models should determine the granularity of compositionality that can be sustained at varying levels of the system-on-chip. The balance between the level of computational complexity supported at local units and the efficiency of distributed processing is best visualized through a spectrum of Pareto-front mapping solutions~\cite{weng2020dsagen}. Considering the synergistic nature of VSA operators, it is also highly desirable to develop frameworks for hardware-aware vector-symbolic architecture search for efficient model-hardware co-development~\cite{saha2023tinyns}.

\subsection{Unified Benchmarking of VSA Hardware/Software Co-Design Flows}
The myriad combinations of VSA algorithms and hardware technologies have aggravated the demand for a systematic evaluation methodology that enables standardized, fair comparisons of different VSA design flows. This gives rise to the importance of developing unified benchmarking suites, akin to benchmarks in conventional machine-learning systems~\cite{reddi2020mlperf}, robotic systems~\cite{krishnan2022roofline,mayoral2023robotperf}, compute-in-memory systems~\cite{peng2019dnn+}, etc. Such benchmarking allows for assessing the performance of various VSA algorithms on the same hardware architecture, and at the same time facilitates the mapping and evaluation of the same algorithm on multiple hardware platforms. A unified benchmark for VSA should include tests for various cognitive tasks, including learning, common-sense reasoning, planning, and decision-making~\cite {srivastava2023beyond}, and it should also allow researchers to investigate tradeoffs in accuracy, energy efficiency, and performance across a wide range of hardware designs (e.g., NVM-based and FPGAs). 
Additionally, given the increasing complexity of VSA algorithms and the cross-stack nature, building a push-button flow with VSA task requirements as input to automatically generate accelerator design is critical~\cite{wan2022robotic}. We envision the agile framework will intelligently search the huge design space and automatically choose the optimal algorithm-hardware parameters with the help of modular kernels, benchmarking, and machine learning-assist methods.

\subsection{Cross-Layer Integration of Heterogeneous NVSA Workloads}
NVSA workloads are heterogeneous due to the varied types of information processing involved. As explained in Section~\ref{sec:codesign}, the diversity observed in such workloads leads to efficiencies that cannot be obtained from single specialized architectures (e.g., DNNs). These efficiencies cannot be fully exploited unless there is a systematic integration mechanism that spans different layers of the computing stack, ranging from hardware to software~\cite{wan2024towards,wan2025cogsys,ibrahim2024special}. In broad terms, NVSA workloads are not compatible with all existing hardware designs; therefore, making progress in this field cannot be achieved unless such a gap is closed---a situation often referred to as \emph{winning the hardware lottery}~\cite{hooker2021hardware}. Particular research areas that need greater attention are threefold: (1) the realization of reconfigurable dataflows that can switch between different cognitive modalities (e.g., neural and symbolic) while achieving the highest compute utilization; (2) the development of a unique memory system that seamlessly interleave different types of cognitive data types (e.g., scalars, tensors, and HD vectors); and (3) defining a ``cognitive'' interconnection protocol that coordinates data movement between units across all the layers. 

The above challenges (and several others) provide ample opportunities for research development at all levels. Exploring innovative solutions to these challenges and embracing interdisciplinary approaches can pave the way for even more breakthroughs in VSAs, significantly advancing cognitive systems on a broader scale.
\section{Conclusion}
\label{sec:conclusion}

This paper surveys the current state of VSA systems and emphasizes the crucial role of hardware/software co-design. We explore diverse VSA algorithms and hardware technologies, and present a new framework bridging the gap between software-level explorations and efficient hardware designs. Discussions on open research challenges and opportunities serve as a call for action on cross-layer collaborations to bring this promising paradigm closer to ubiquitous applications. 

\bibliographystyle{plain}
{
\hyphenpenalty=10000
\exhyphenpenalty=10000
\sloppy
\bibliography{main}
}

\vfill

\end{document}